\documentclass[a4paper,12pt]{article}
\usepackage[utf8]{inputenc}
\usepackage{rotating}
\usepackage{booktabs}
\usepackage{makecell}
\usepackage{multirow}
\usepackage{amssymb}
\usepackage{lipsum}  
\usepackage{ragged2e}
\usepackage{threeparttable}
\usepackage[margin=2cm]{geometry}
\usepackage{tabularx}
\usepackage{booktabs}
\newcolumntype{Y}{>{\raggedleft\arraybackslash}X}
\usepackage{subcaption}
\usepackage{flafter} 
\usepackage{array}
\usepackage[section]{placeins}
\newcolumntype{C}{>{$\displaystyle}c<{$}}

\usepackage[utf8]{inputenc}
\usepackage{geometry}
\usepackage{setspace,pdflscape}
\usepackage{xcolor}
\usepackage{hyperref}
\hypersetup{
    colorlinks=true,
    linkcolor=black,
    menucolor=black,
    filecolor=black,      
    urlcolor=blue,
    citecolor=blue,
}
\usepackage{lipsum}
\usepackage{float}
\usepackage{abstract}
\usepackage[version=3]{mhchem} 
\usepackage{natbib}
\usepackage{cite}
\usepackage{graphicx}
\usepackage{graphics}
\graphicspath{ {figures/} }
\usepackage{blindtext}
\usepackage{setspace}
\usepackage{booktabs}
\usepackage{array}
\usepackage{amsmath} 
\usepackage{caption}
\usepackage{titling,lipsum}
\usepackage{tikz,lipsum,lmodern}
\usepackage[most]{tcolorbox}
\usepackage{tikz}
\usetikzlibrary{arrows,positioning} 
\usepackage{flowchart}
\usetikzlibrary{arrows}

\title{Procompetitive effects of vertical takeovers. \\ Evidence from the European Union\thanks{\scriptsize{We are grateful for the comments and suggestions received by Davin Chor, Basile Grassi, Sergei Guriev, Gianmarco Ottaviano and Maurizio Zanardi. We are also thankful for helpful discussions with conference participants at ASSA 2023, ETSG 2022, JEI 2022, RES Annual Conference 2022, FIW Research Conference 'International Economics' 2022, and ETSG 2021.  Armando Rungi claims financial support from the PRIN project 2022W2245M, financed by the European Union - Next Generation EU.
}}}

\author{Chiara Bellucci\thanks{\scriptsize Mail to: chiara.bellucci@imtlucca.it. Laboratory for the Analysis of Complex Economic Systems, IMT School for Advanced Studies, Piazza San Francesco 19 - 55100 Lucca, Italy.}\and Armando Rungi\thanks{\scriptsize Mail to: armando.rungi@imtlucca.it. Laboratory for the Analysis of Complex Economic Systems, IMT School for Advanced Studies, Piazza San Francesco 19 - 55100 Lucca, Italy.}}

\date{This version: May 2025}

\begin{document}
 \begin{titlingpage}

\maketitle
\setcounter{page}{1}
\singlespacing
\vspace{-5mm}

\begin{abstract}
 
This study investigates the causal impact of takeovers on firm-level financial accounts on a sample of 4,482 targets in the European Union in the period 2007- 2021. Findings suggest that horizontal integrations do not have a statistically significant impact, while vertical takeovers bring about a lower markup (0.7\%), a larger market share (2.5\%), a higher profitability (2.3\%), and a lower capital intensity (7.2\%). The impact of vertical integrations grows over time, and it is higher when the corporate perimeter of the acquirer is bigger. Our results point to strategies aimed at eliminating double profit margins along supply chains. Finally, we reconnect with the debate initiated by the U.S. Vertical Merger Guidelines in 2020 and 2023, where the presumption of harm after vertical deals has been softened, thus considering procompetitive effects, but the discussion of potential foreclosure risks has been expanded.

\medskip



\end{abstract}
\end{titlingpage}

\newpage

\section{Introduction}\label{sec: intro}

Recent years have seen a renewed debate about the role of vertical acquisitions as a consequence of their growing importance in competitive markets. According to the Institute for Mergers, Acquisitions and Alliances (IMAA), the number of deals was 5,009 in 1990 for a value of about 203 billion dollars, but it steadily increased over time, peaking at 58,308 in 2021, reaching a value of about 5.3 trillion dollars. Unlike horizontal acquisitions, a vertical acquisition does not eliminate direct competition. In a vertical acquisition, a target company is taken over by one of its buyers or suppliers along the supply chain, and competition authorities have often considered these cases more favourably. Yet, the attitude has changed. In the USA, the Federal Trade Commission reviewed the 2020 Vertical Merger Guidelines because they included a \textit{‘flawed discussion of the purported procompetitive benefits’} \citet{USguidelines2020, USguidelines2023}. Similarly, in the United Kingdom, the Competition and Markets Authority (CMA) no longer describes vertical mergers as \textit{'generally benign'} \citet{UK2021}. In the European Union, there has been no official change since the latest non-horizontal merger guidelines in 2008 \citet{EU2008}, but case practice continues to evolve \citet{Karlinger_et_al_2020}.

We contribute with an empirical study to test the impact of horizontal and vertical takeovers on a dashboard of firm-level outcomes in the European Union from 2007 to 2021. For our purpose, we exploit a complete and representative sample of about 579,354 manufacturing firms active in the European Union in 2007-2021 to estimate production functions and derive both markups and total factor productivities (TFP) following standard methodologies proposed by \citet{DeLoecker_Warzynski_2012} and \citet{Ackerberg_Caves_Frazer_2015}. Then, we show a few stylized facts where targets of takeovers are, on average, bigger, more capital-intensive, and have a higher market share. Interestingly, targets also average lower markups and TFPs. The latter evidence is puzzling because one would expect that integration entails a higher market power and more efficiency. Therefore, motivated by preliminary evidence, we question simple correlations to design an empirical strategy that challenges endogeneity and identifies the direction of causality. Since acquiring firms can screen targets and pick the ones with better performance, we want to compare targeted firms with counterfactual observations with similar economic potential. We propose to look at a dashboard of financial accounts that allows us to qualify the impact of the takeover (market share, sales, variable costs, productivity, profitability, capital intensity, and financial constraints).

Our identification strategy combines a difference-in-difference specification with a propensity score matching exercise to control for the endogenous selection of targeted firms based on observable financial information. The aim is to consider cherry-picking when parent companies acquire control over firms after anticipating their market potential. Following most recent developments by \citet{Callaway_Santanna_2021}, our specification is robust to the presence of staggered treatments (in our case, the takeovers) that can occur in different periods, i.e. when cohorts of acquisitions distribute unevenly over time. We run a doubly robust estimator \citep{Sant'Anna_Zhao_2020} to identify the counterfactual of firms that were not targets of takeovers by a matching procedure with endogenous covariates and inverse probability weights. The parallel trend assumption holds after conditioning on \textit{ex-ante} firms' observable characteristics.

Interestingly, we don't find any statistically significant increase in markups or market shares when we pool all the targets together. However, companies pursue distinct strategies with acquisitions. Through horizontal integration, a company acquires a competitor to consolidate market share. Alternatively, a firm may opt for vertical integration by acquiring a company along the supply chain — either downstream among buyers or upstream among suppliers — to secure lower-cost or higher-quality intermediate inputs, thereby gaining an indirect competitive advantage over rivals. We therefore separate vertical from horizontal takeover strategies and find that, on average, lower markups (-0.7\%) are charged after vertical integrations. In this case, associated with lower markups, we find that the same targets also benefit from increased market shares (2.5\%) and their scale of operations increases, since both sales and variable costs proportionally rise by 2.9\% and 3.2\%, respectively. Additionally, we record a weakly significant positive impact on productivity, which is, however, offset by a negative impact on financial constraints, possibly due to the costs of the deal that the target company is usually induced to share with the acquiring company. Notably, when looking at event studies, we find that the impact on markups after vertical integrations is higher with time passing, until reaching a peak of about 4\% in the last year in our analysis period. Moreover, firms' sales, variable costs and profitability are similarly observed to increase with time, reaching yearly peaks at 11.5\%, 12\%, and 18.5\%, respectively, after twelve years from the takeover.

Our findings suggest that vertical integration along supply chains enhances efficiency across different dimensions. First, firms engaging in vertical takeovers can achieve both economies of scale and scope, as indicated by the proportional increase in sales and variable costs, while also eliminating redundant facilities like head offices, as reflected in the significant decrease in capital intensity. Additionally, they can facilitate the transfer of intangibles such as know-how and human capital, though these effects remain challenging to capture empirically. Second, our results suggest that vertical integration reduces double (or multiple) profit margins by consolidating buyers and suppliers under a single headquarters, thereby lowering successive markups along the supply chain. After becoming part of the same corporate entity, it is possible for buyers and suppliers together to increase market efficiency and charge lower final prices, thus pointing to increasing market shares and sales. From a more general perspective, we argue that vertical integration strategies can yield overall efficiency gains while sustaining volume growth and they can contribute to reducing overall welfare inefficiencies by internalizing part of the production processes. In addition, we find that the magnitude of a markup reduction is higher when the number of previously integrated subsidiaries is higher. From our point of view, the latter evidence indicates that there is more scope for eliminating double margins, hence lowering markups, when there is a bigger perimeter of the supply chain that the parent company already coordinates.

Please note that, according to our findings, markup reductions can occur whether the target firm is upstream or downstream of the parent company. When the target company operates upstream, it can sell intermediate inputs at a lower price without affecting marginal costs, thereby reducing markups and enabling the parent company to internalize part of its profits. Additionally, markup reductions may result from a combination of lower selling prices — supported by the significant increase in sales — and higher marginal costs, proxied by variable costs, driven by increased input usage as production scales up. If the target company is downstream, a reduction in its markup may again stem from both price reductions and rising marginal costs. In our financial accounts dashboard, we verify that, on average, there is no asymmetry between sales and variable costs, as their ratio remains statistically unchanged after the takeover.

When we introduce sensitivity checks, we find that markup reductions are mainly found in medium- and high-tech industries, where there is a bigger scope for price reduction. Besides industry heterogeneity, we find that the geographic sample composition also matters because markup reductions are relatively higher when we consider old Member States of the European Union and, in particular, the top 5 biggest economies in the period we consider. Eventually, a relatively higher markup reduction is detected in the case of foreign takeovers, which represent about one-third of the entire sample.

Finally, we reconnect with the debate on the health of competition policies. On the one hand, the lack of a systematic impact on European markups or market shares after horizontal takeovers seems to be the result of a good competition framework on the old continent. Takeovers have been largely acknowledged as a fundamental channel through which markets can concentrate, and that is why they have always been under the scrutiny of competition authorities. In the European Union, cases of mergers and takeovers fall under the European Competition Law to preserve the benefits of the Single Market. Under the European Union Merger Regulation (EUMR), Art. 2(3), for a merger to be declared compatible with the Single Market, it must not create or strengthen a dominant position. Therefore, there is a general acknowledgement that the intention of the regulators has been to establish a way first to prevent and then to sanction the emergence of dominant positions.

On the other hand, our results point to pro-competitive effects for vertical takeovers, in contrast with the lack of evidence in single competition cases \citep{Dewulf_et_al_2022}. Yet, despite our evidence of direct welfare efficiency gains, further work is needed to understand whether foreclosure effects on inputs' markets can offset the benefits from vertical integration. From this perspective, we welcome the most recent U.S. Vertical Merger Guidelines of 2023, which have expanded the economic discussion to encompass potential foreclosure risks after vertical deals. 

The remainder of the paper is organized as follows. The next section \ref{sec: lit} relates our contribution to previous scholarly literature. Section \ref{sec: data} describes our data structure and provides preliminary evidence on the evolution of markups and other economic variables of interest. Section \ref{sec: strategy} describes the identification strategy to derive the impact of takeovers on market power and other firms' dimensions. Section \ref{sec: robustness} controls for the robustness and sensitivity of our findings. Finally, Section \ref{sec: conclusions} concludes.

\section{Related literature}\label{sec: lit}

Our contribution broadly relates to recent works investigating market power. Several authors detect a rise in global market power using markups derived from production functions \citep{Hall_2018, DeLoecker_Eeckhout_2020, Diez_at_al2021, Bighelli_2023}. Yet, evidence for the European Union is mixed. \citet{Bighelli_2023} show that firm concentration has increased in Europe in the last decade. At the same time, they find a positive and significant correlation between rising sector-level concentration and increases in sector-level productivity, which can still benefit the consumers. Differently, \citet{McAdam_et_al_2019} find that concentration ratios in the euro area have remained broadly flat in the last ten years, thus suggesting that competition intensity may have been reasonably stable, while markups have declined marginally since the late 1990s. Pooled estimates at the world level \citep{DeLoecker_Eeckhout_2018} report a stable increase in global markups, even though it is reasonable to expect a certain degree of heterogeneity among different countries and markets. When it comes to explaining the trends, \citet{DeLoecker_Eeckhout_2020} noticed for the U.S. that the upper tail of the distribution mainly drives the rise in markups. Market shares are reallocated towards \textit{superstar firms} with higher markups and lower labour shares \citep{VanReenen_2018, Autor_et_al_2020, alviarez_2020}. The latter emerged thanks to newly available technologies, declining trade costs and the fall of non-tariff barriers enabled by globalisation and deep regional integration agreements. In this sense, the general idea is that markups are a possible threat to competitive markets and business dynamism, resulting in lower levels of social welfare through a misallocation of productive resources \citep{baqaee_2020} and possibly lower labour shares \citep{deb_2022}. 

Against the previous background, takeovers are considered a channel that can drive changes in market power. On the one hand, takeovers can increase prices at the expense of consumers; on the other hand, productivity gains due to knowledge transfer, lower marginal costs of cheaper intermediate inputs and the reallocation of resources to more efficient uses may benefit consumers in the form of improved products or lower prices. Recent empirical studies have found contrasting results about the final impact of M\&A activities on market power, concentration and productivity. \citet{Stiebale_Vencappa_2018} find that acquisitions in India are associated with increased quantities and markups but with lower marginal costs. \citet{Blonigen_2016} use U.S. Census Bureau data on manufacturing plants to find significant increases in average markups from M\&A activity but little evidence for productivity gains. Also, \citet{McGukin_Nguyen_1995}, \citet{Gugler_et_al_2003}, and \citet{Maksimovic_et_al_2011} rely on firm-level data to estimate the impact of firms' acquisitions on market power and productivity and find evidence of a positive effect on productivity measures. 

Yet, firms may engage in different M\&A strategies depending on the goal they want to achieve. Changes in market power after the acquisition may occur due to horizontal integration when a market player absorbs a direct competitor and adds market shares and profits. In this case, the firm aims to increase its market share and achieve economies of scale to increase profits. Although many empirical studies find evidence of increased market power after acquisitions, others, such as \citet{Bertrand_Zitouna_2008}, do not find significant changes in profit margins after horizontal integration. \\ When one company takes over a customer or a supplier, as in the case of vertical integration, the efficiency gains can span across different dimensions. Many empirical works have found positive evidence of such efficiencies by looking at effects on prices, cost, investments and product quality, as summarised by \citet{lafontaine_2007} and more recently by \citet{slade_2021}.
Other works have found that half of the firms that vertically integrate do not exchange goods, but intangible inputs \citep{atalay_2014}, and eventually obtain gains achieved through more efficient use of, for example, technology and logistics \citep{Hortaccsu_Syverson_2007}. 

Other empirical studies examining specific markets provide evidence of efficiency gains from vertical integration. For instance, \citet{ciliberto_2006} shows that integrating doctors and hospitals increases investment in healthcare services. Similarly, \citet{chipty_2001} and \citet{crawford_2018} find that vertical integration can, though not necessarily, increase the coverage of certain types of TV channels. \citet{gil_warz_2015} further document that vertical integration is associated with higher-quality video games.

Our paper contributes specifically to the literature on efficiency gains stemming from the elimination of double marginalisation (EDM), a concept first formalised by \citet{Spengler_1950}. Since then, EDM has played a central role in antitrust discussions\footnote{See \citet{baker_2018}, \citet{salop_2018}, and \citet{kwoka_2019} for a discussion on vertical mergers from an antitrust perspective.}, and its efficiency implications remain actively debated. While several empirical studies find that EDM following vertical integration leads to price reductions \citep{chipty_2001, Gil_2015, crawford_2018}\footnote{Recent work focus on estimating double markups \citep{duran_2022} and markups along supply chains, highlighting how backward integration can lower input costs through improved access or buyer power \citep{morlacco_2019, rubens_2023}, while forward integration may reduce average costs via economies of scale \citep{antras_2020}.}, other evidence suggests that EDM does not always generate efficiency gains. For example, in the presence of multiproduct firms, prices of non-integrated products may increase \citep{salinger_1991, luco_2020}, and cost efficiencies may not be fully passed through to consumers due to market concentration \citep{grieco_2018}.

Other consequences of vertical integration may be the foreclosure of other competitors \citep{Spengler_1950,  comanor_1967, hastings_2005, Berto_2007}. Eventually, the overall welfare effects from the elimination of double margins are ambiguous, as pointed out by \citet{chone_2023}, because they depend on the distribution of bargaining power in upstream and downstream markets, possibly bringing heterogeneous impacts on the ability to source from other independent suppliers.

In this regard, when we discuss our findings, we also relate to results from seminal industrial organisation literature \citep{berry_2019}, according to which higher markups and market concentration \textit{per se} do not imply lower social welfare. The heterogeneity of market structures across industries can offer differing explanations for rising markups, such as in the case of increasing endogenous fixed costs that could be associated with lower marginal costs. In the case of technology-intensive industries, the reliance on R\&D efforts is higher than in lower-tech industries. Our work tests that results are heterogeneous by technological trajectories, and medium-high tech industries, on average, respond more to the elimination of double margins after vertical integrations.

Finally, this contribution also relates to previous works showing how different institutional settings in the EU and the U.S., including antitrust and regulation by competition authorities, may lead to varying patterns of market power across countries. In this perspective, a historical decline in antitrust enforcement in the United States has led to harmful market concentration and increased prices and barriers \citep{Grullon_et_al_2019}. In contrast, European markets have become more competitive due to stronger and more independent enforcement \citep{Gutierrez_Philippon_2023}.

\section{Data and Preliminary Evidence}\label{sec: data}

\subsection{Data description}

We source firm-level financial accounts and ownership information from the Orbis database compiled by the Bureau Van Dijk\footnote{The Orbis database standardises firm-level financial accounts and ownership on a global scale. It also includes an ownership module that tracks changing shareholding information at the firm level. Orbis data have been increasingly used for firm-level studies on multinational enterprises. See for example \citet{Cravino_Levchenko_2016}, \citet{DelPrete_Rungi_2017}, \citet{DelPrete_Rungi_2020}, \citet{alviarez_2020}}. First, we collect complete financial information on 579,354 European firms in the manufacturing industries active from 2007 to 2021. Following international standards \citet{oecd2005mne, eurostat2007for, unctad2009fdi, unctad2016inv}, we define a subsidiary as a company with a corporate controlling shareholder with a direct or indirect absolute majority (50\%+1) of voting rights at the shareholder assembly. Therefore, we can define a takeover as a change in the controlling majority when a new controlling corporate shareholder emerges in our observation period.

\begin{figure}[H]
\centering
\caption{Distribution of markups in the European Union}
\centering
\begin{minipage}{0.55\textwidth}
\includegraphics[scale=0.12]{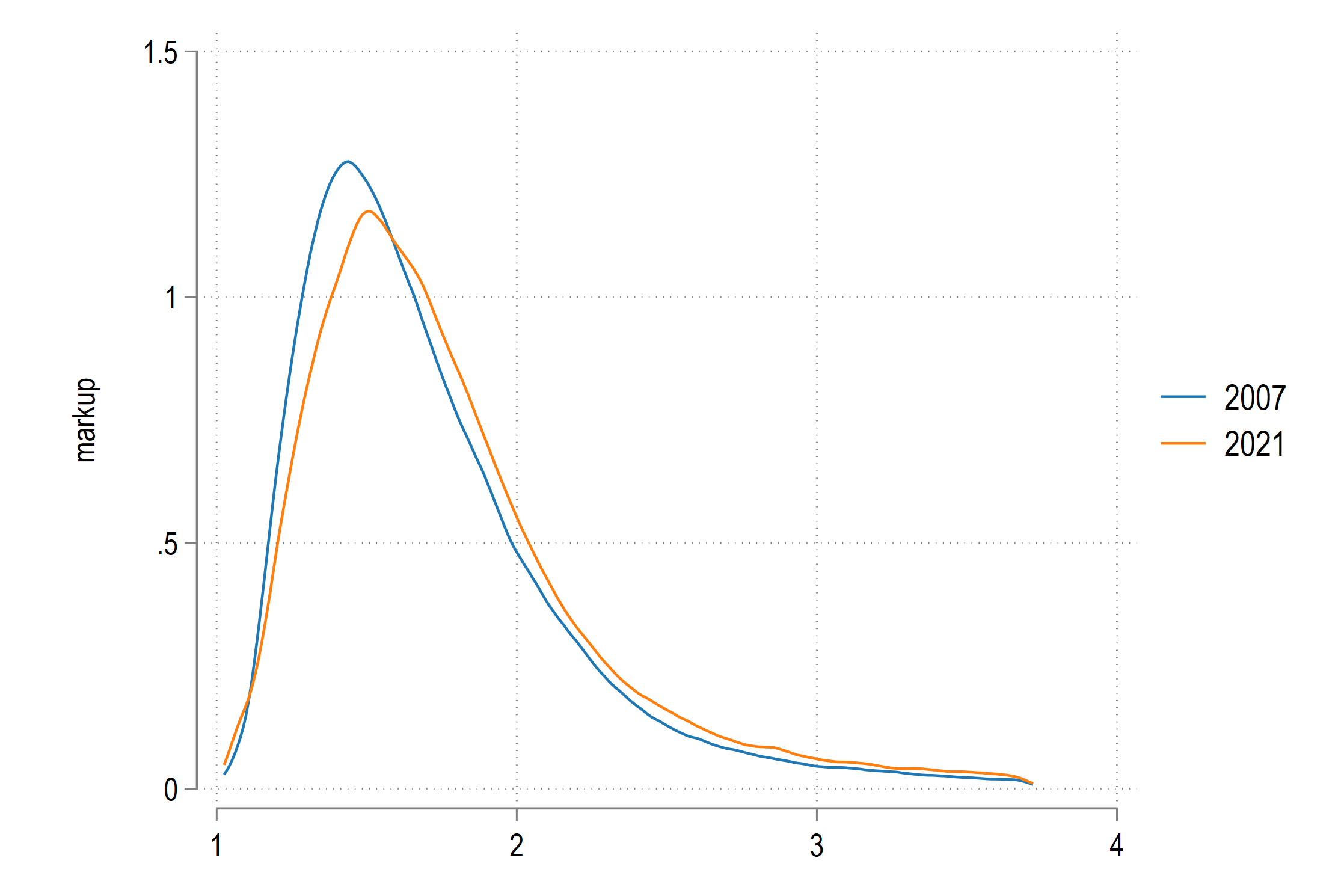}
\footnotesize{Note. Distribution of markups of European manufacturing firms in 2007 and 2021. Markups are estimated following \citet{DeLoecker_Warzynski_2012}.} The distribution presents a mean value of 1.74 with a median equal to 1.63 and a standard deviation of 0.47.
\end{minipage}
\label{fig:fig1}
\end{figure}

For the scope of our study, we estimate production functions to derive firm-level measures of markups and Total Factor Productivity (TFP) following the methodologies proposed by \citet{DeLoecker_Warzynski_2012} and \citet{Ackerberg_Caves_Frazer_2015}\footnote{In Appendix A, we describe the details of the procedure for deriving markups, and coefficients of sector-level production functions are reported in Appendix Table \ref{tab:profuction_function}. Importantly, in Section \ref{sec: robustness}, we address methodological concerns raised by previous literature. Crucially, \citet{Basile_et_al_2024} show that revenue-based markup estimates are equal to true markups up to a constant; thus, their correlation is equal to one. This is true in the case of constant and non-constant elasticities. The constant simultaneity bias does not fade away; it affects magnitudes, and we can still discuss rankings, variances, and growth rates.}. Usefully, in Figure \ref{fig:fig1}, we show the distribution of markups we obtained for all firms in 2007 and 2021, respectively. In line with previous studies, most firms have relatively low markups, while only a few firms on the right tail have disproportionately higher market power. We observe a slight shift in the distribution of markups at the end of the period, suggesting that markups have slightly increased over time.

\begin{figure}[H]
\centering
\caption{Evolution of aggregate markups }
\centering
\begin{minipage}{0.55\textwidth}
\includegraphics[scale=0.11]{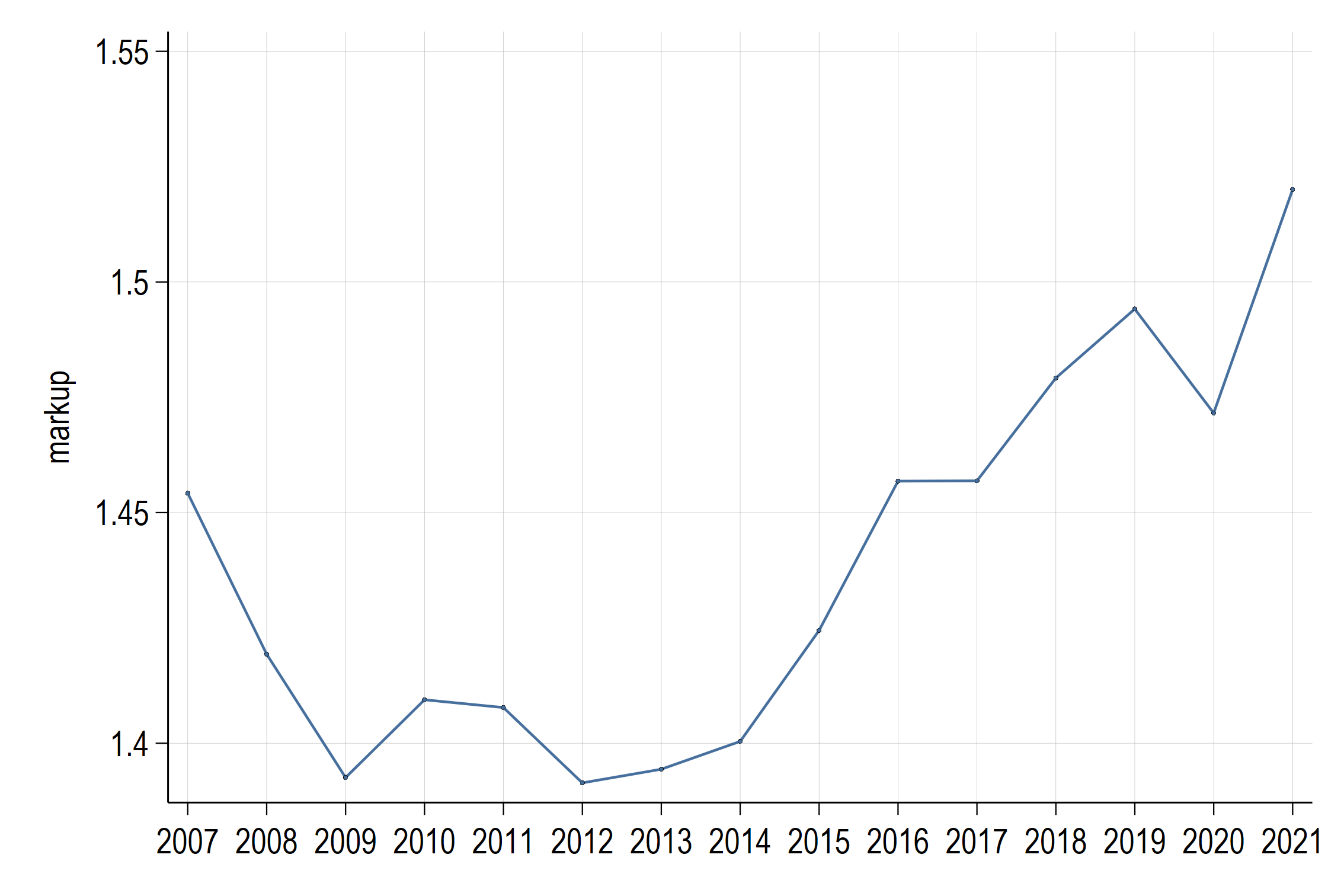}
\footnotesize{Note. The figure reports European manufacturing firms' sales-weighted average markups in 2007-2021. Markups are estimated following \citet{DeLoecker_Warzynski_2012}.}
\end{minipage}
\label{fig:fig2}
\end{figure}

To provide further evidence about changing patterns, we aggregate sales-weighted markups in Figure \ref{fig:fig2}. Even though the time span covered in our analysis is insufficient to provide a long-term trend, we can fairly notice that markups are volatile, albeit generally increasing from 2012 to 2021, after a period of decrease during the financial crisis. Yet, aggregate estimates might hinder the emerging heterogeneity when considering different industries. For this reason, we plot separate trends by 2-digit NACE industries in Appendix Figure \ref{fig:fig3}. Despite the great degree of heterogeneity in average markups across sectors, we record an overall increasing trend over time \footnote{This evidence is confirmed by a specific exercise to identify the linear trend in markups after we run a simple least-squares, where the dependent variable is markups, the main regressor is a time trend, and we control for firm-level covariates including size, capital intensity, market share and labor productivity. We also include fixed effects and clusters of standard errors at the NACE 3-digit level. Eventually, we find that markups have been increasing overall linearly by a non-negligible 0.2\% per year in our analysis period.} with a few exceptions, such as the manufacturing of food and textiles products as well as wood, transport equipment, and furniture.

Taken together, descriptive evidence confirms an increasing trend of market power. In the remainder of the chapter, we shed light on the peculiar role of M\&A activities in affecting the trends of markups and other firms' outcomes. In particular, in the empirical analysis, on top of markups, we will look at a full dashboard with different firms' outcomes that can help understand the overall impact of takeovers: market shares, sales, variable costs, productivity, return on investments, capital intensity, liquidity, and solvency ratios.
It is reasonable that firms, after a takeover, will experience changes in their business strategy and organizational structure that will likely affect the overall performance. Thus, we first look at market shares to check whether there is a direct effect, especially in cases of horizontal integration. We analyze the impact on sales and variable costs as proxies of the firm's growth volume. We measure productivity by looking at Total Factor Productivity (TFP), estimated following \citet{Ackerberg_Caves_Frazer_2015}. We also analyze the impact of takeovers on return on investments, measured as the ratio between profits and lagged fixed assets and capital intensity. In conjunction with markup changes, return on investments can increase or decrease due to a combination of changes in prices and volume of activity. Also, capital intensity, measured as fixed assets per employee, can be impacted due to synergies resulting from rationalizing the production process. Appendix Table \ref{tab:description} shows summary statistics for the variables included in the analysis, while Appendix Table \ref{tab:corr} shows pairwise correlations between variables. 

For our purpose, we extract from our general sample a total of 4,482 cases of firm-level takeovers, whose coverage is reported in Appendix Table \ref{tab:year_distribution}. Please note that we exclude cases of multiple acquisitions of the same subsidiary, assuming that treatment can occur at most once for each firm. This is consistent with the idea that direct investment has a longer-term perspective, and thus, any shorter-run change in equity in an investor's portfolio is not able to significantly impact the management of economic activities. In Appendix Table \ref{tab:sector_distribution}, we have a look at the sample coverage of takeovers across sectors.\\

\subsection{Targets and acquirers}

In these paragraphs, we provide descriptive evidence on both targets of takeovers and their acquirers. We begin by looking at how firms that have been taken over compare with other sample firms in Table \ref{tab:descriptives}. We perform t-tests for a set of variables of interest to check whether there is any systematic difference across the two subsets. Indeed, we acknowledge that the average values of sales, capital intensity, fixed assets, added value, number of employees, market shares and variable costs are higher in the case of firms that have been acquired vs the ones that never changed ownership majorities. On the other hand, we find lower average values for markups and productivity for targeted firms, but no significant differences in profitability. 

From another perspective, we can say that it is very likely that bigger and more productive firms are more attractive targets for acquisitions. From this point of view, it is clear that differences in firms' performances are endogenously related to events of acquisitions. Therefore, our empirical strategy will take care of the endogenous selection by acquiring firms and, thus, challenge reverse causality to establish the causal contribution of takeovers to firm-level outcomes.

\begin{table}[htbp]
\small
\caption {Targeted firms vs. non-targeted firms} \label{tab:descriptives}
\begin{center}
\begin{threeparttable}
\begin{tabular}{lrrr}
\toprule
Variable	&	Average target firm	&	Average non-target firm	&	t-test $\Delta \neq0$	\\
\hline							
Markup 	&	1.54	& 	1.65	&	-0.11***	\\
Sales	&	 31,100,000 	& 	8,299,760	&	22,800,000***	\\
ROI	&	6.40	&	11.17	&	-4.77	\\
Capital intensity	&	97,166	&	 61,255	&	 35,911***	\\
(log of) TFP	&	9.73	&	9.76	&	-0.31***	\\
Fixed assets	&	13,500,000	&	4,181,983	&	9,350,618***	\\
Added value	&	8,373,059	&	2,621,432	&	5,751,627***	\\
N. of employees	&	118	&	41	&	77***	\\
Market share	&	0.003	&	0.0008	&	0.002***	\\
Variable costs	&	22,300,000	&	5,940,373 	&	16,300,000***	\\

\bottomrule
\end{tabular}
\begin{tablenotes}[para,flushleft]
\footnotesize{The table reports average values of variables of interest with a t-test for significance. Markups are estimated following \citet{DeLoecker_Warzynski_2012}. *** stands for $p<0.001$.}
\end{tablenotes}
\end{threeparttable}
\end{center}
\end{table}

\begin{figure}[H]
\caption{Heterogeneity in the corporate perimeter}
\centerline{\includegraphics[scale=0.2]{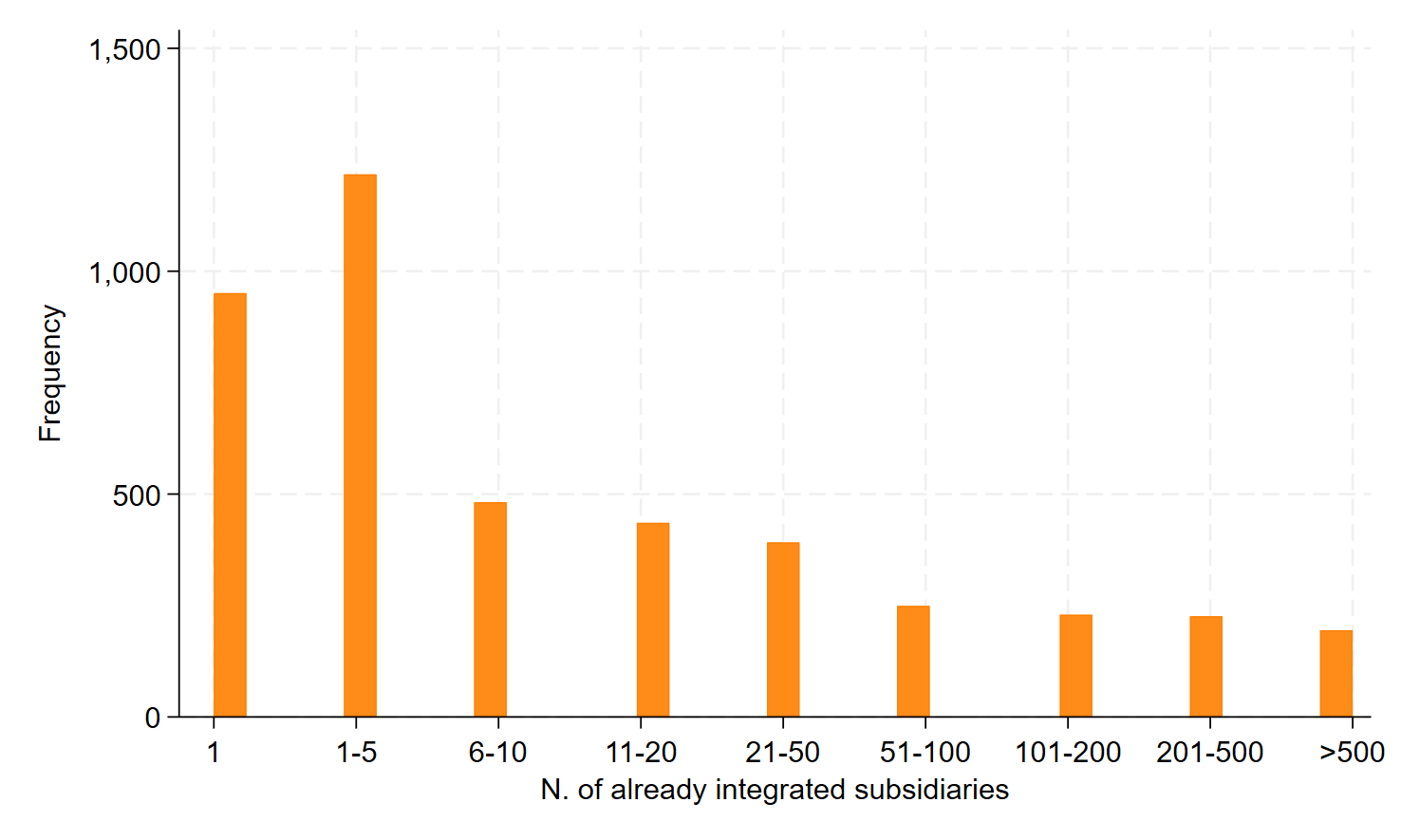}}
 \begin{tablenotes}
      \footnotesize
      \singlespacing
      \item Note: The figure shows the corporate perimeter, that is, the number of subsidiaries previously integrated by the parent company before the new takeover in our period of analysis (2007-2021). Subsidiaries can be directly or indirectly owned with a majority equity stake.
   \end{tablenotes}
\label{fig: corporate_perimeter}
\end{figure}

Before proceeding to causal inference, we believe it is also important to look at the other side of the takeover deals. In our sample, we have a total of 3,776 acquiring parent companies. Crucially, many of them already control other subsidiaries before they complete the takeovers we investigate in our sample. That is, they have a corporate perimeter that encompasses the parent company itself and the subsidiaries controlled on the date before the takeover is completed. In Figure \ref{fig: corporate_perimeter}, we describe the distribution of such corporate perimeters, thus representing the heterogeneity in the segments of supply chains already integrated. Interestingly, we have only 21.7\% of the acquiring companies (821) that were independent, and the corporate perimeter is equal to one because it includes only headquarters. Notably, there are about 27.8\% of sample acquirers that controlled up to six other subsidiaries, while a relatively small bunch of parent companies, 4.5\% of the sample, controlled bigger networks with more than 500 subsidiaries before engaging in yet another takeover.

\begin{figure}[H]
\caption{Markups of targets and corporate perimeters of the acquiring parent company}
\centerline{\includegraphics[scale=0.28]{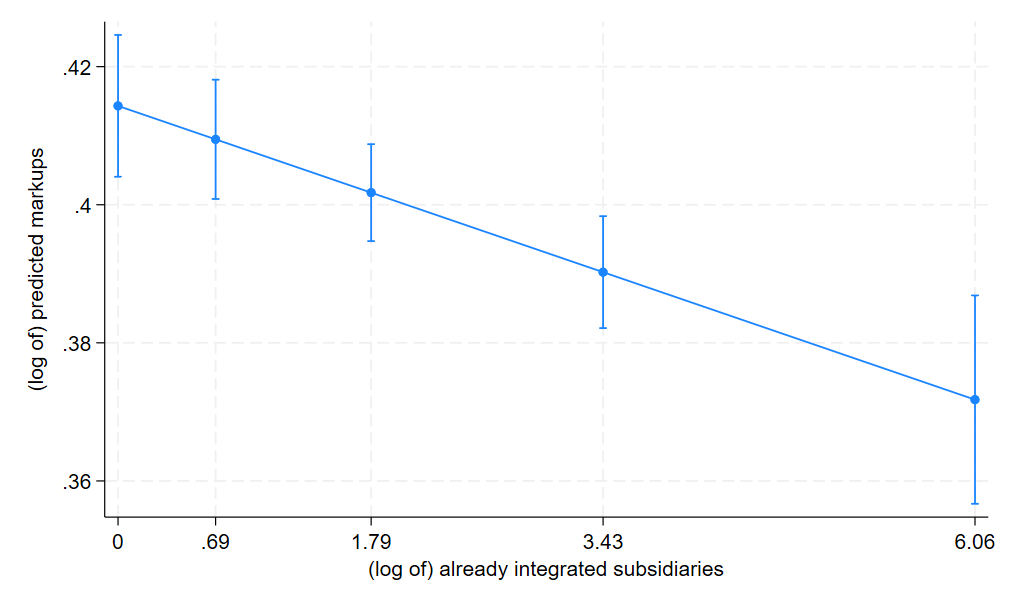}}
 \begin{tablenotes}
      \footnotesize
      \singlespacing
      \item Note: The figure shows linear predictions after a basic regression where the dependent variable is the (log of) markup of targeted companies in our period of analysis (2007-2021), and the regressor is the number of already integrated subsidiaries by the acquiring parent company. The confidence interval is set at 95\%.
   \end{tablenotes}
\label{fig: association markups and the corporate perimeter}
\end{figure}

Thus, the question arises: What should we expect when bigger corporations complete a new takeover? Will the market power of the new subsidiaries be higher or lower? In Figure \ref{fig: corporate_perimeter}, we report a simple stylized fact. Interestingly, we find a negative association between the (log of) markups of targeted companies and the (log of) number of companies already present in the corporate perimeter. More specifically, after we run a simple least-square regression, we draw linear predictions at different percentiles of the distribution of already integrated subsidiaries, and we report graphical evidence of predictive margins in Figure \ref{fig: association markups and the corporate perimeter}. The predictive margins' plot indicates that lower markups are detected when acquirers have bigger corporate perimeters. Originally, this is highly counterintuitive, as we expected that takeovers were a channel to exploit market power. We expected an opposite sign in the relationship: higher markups in relation to bigger corporate perimeters. 

In the following paragraphs, we will specifically address causality concerns, and we will discuss the credibility of the latter stylized fact.

\section{Empirical Strategy}\label{sec: strategy}

This section tests the causal impact of takeovers on firm-level outcomes. We implement an empirical strategy in two steps. First, we combine a propensity score matching with a difference-in-difference model with a panel data setting when staggered treatments occur in multiple periods. For our exercise, we rely on the procedure proposed by \citet{Callaway_Santanna_2021}. We consider as treated those firms that new parent companies took over, compared to a control group obtained after a propensity score matching. In this case, \citet{Callaway_Santanna_2021} propose a matching procedure that exploits all available information on untreated companies with the adoption of inverse probability of treatment weights. The scope is to eliminate the endogenous selection bias of targeted firms into the treatment since we assume that acquirers can screen the firms with the best economic potential before a bid. After the matching, our preferred methodology improves on a classical difference-in-difference approach because it considers the bias of heterogeneity in treatment timing, i.e. when takeovers can occur endogenously and asymmetrically over the timeline we can observe.

In the second step of our analysis, we separate events of vertical integration from the rest of the takeovers, as the former indicates an organisation of supply chains within or across national borders. The intuition is that the vertical integration of supply chains under the coordinated management of a parent company implies a different organisation of production processes, where an intra-firm shipment of intermediates can occur, whose impact on market power has been relatively neglected by previous scholarly literature.

\subsection{Market power and takeovers}
First, we estimate firm-level markups using the well-established methodology proposed by \citet{DeLoecker_Warzynski_2012}. Please refer to Appendix A for a detailed description of how to recover firm-level markups using the production function approach. We address some of this estimation method's main limitations and concerns in Section \ref{sec: robustness}. Briefly, we believe that our preferred methodology allows us to obtain wider datasets with firms across many industries and countries at the cost of simplifying assumptions on underlying market structures \citep{berry_2019}.

To estimate the causal impact of firms' acquisitions, we follow the difference-in-difference strategy proposed by \citet{Callaway_Santanna_2021} in a panel setting, since: i) takeovers can occur in multiple periods; ii) we have variation in treatment timing, as we observe an increasing trend in takeovers; iii) we can assume that the parallel trends assumption holds only after conditioning on observed firm-level characteristics. \\

Our doubly robust estimator identifies multiple $ATET(g,t)$ for each treated firm cohort. Each cohort represents a group $g$ of firms taken over in the same year $t$. It is possible to estimate a set of coefficients, one for each cohort, to track down the impact of the takeover over time. Thus, one can aggregate and obtain a unique coefficient that aggregates the impact of takeovers over the entire timeline. The estimator is obtained as follows:

\bigskip 
\begin{equation}\label{ate}
ATET(g,t) = \mathbb{E} \left[ \left( \frac{G_g}{\mathbb{E}{[G_g]}} -\frac{\frac{p_g(X)C}{1-p_g(X)}}{\mathbb{E}{\left[\frac{p_g(X)C}{1-p_g(X)}\right]}} \right) \left( Y_t - Y_{g-1} - m_{g,t}(X) \right)\right]
\end{equation}

where $G_g$ is a binary variable that is equal to 1 if a unit is first treated in period $g$ and $C$ is a binary variable equal to 1 for firms that have never been a target, or they haven't been a target yet\footnote{Please note the peculiarity of this methodology that allows to include in the control group firms that have not been acquired yet. We believe this further attenuates the selection bias, thus properly considering the panel data dimension of the sample takeovers.}; $p_g(X) = P(G_g = 1|X, G_g + C = 1)$ is the probability of being acquired for the first time in the period $g$ conditional on observed financial information and either being a member of group $g$ or not being acquired in any time period; $ m_{g,t}(X) = \mathbb{E}[Y_t - Y_{g-1} | X, C=1]$ is the population outcome regression for the control group of firms that have never been acquired. We refer to \citet{Callaway_Santanna_2021} for a more complete discussion on the methodology. We choose to use the doubly robust alternative as it provides for a combination of inverse probability weights with an outcome regression approach\footnote{Please note that the IPW and related P-scores are estimated separately for each cohort ($ATT_{g,t}$).} That is, the counterfactual group is obtained by using information about all units that are untreated, assigning to each unit an inverse probability weight of being similar to one that is actually being treated. 
By estimating separate $ATET(g,t)$, we can, therefore, identify differences in the causal effect of the treatment for each cohort, and we are therefore able to determine the degree of heterogeneity of the treatment across groups over time. To estimate the aggregate effect of firms' takeovers on markups, we can finally compute a weighted average of previously defined $ATET(g,t)$ in the following way:

\begin{equation}\label{eq: aggregate_ATT}
    \theta_s^O = \sum_{t=2}^T \theta_s (g) P(G=g)
\end{equation}

where,
\begin{equation}
    \theta_s (g) = \frac{1}{T-g+1} \sum_{t=2}^T \mathbf{1}\{ g \leq t\} ATET(g,t)
\end{equation} 

and $T$ denotes the number of years. $\theta_s (g)$ allows the highlighting of treatment effect heterogeneity with respect to the year in which the firm was acquired. We can aggregate the latter parameter at a higher level and get $\theta_s^O$, which is the overall estimate of the impact of takeovers on firms' outcomes. In other words, the aggregate coefficient is computed as a weighted average of the time-specific parameters $\theta_s (g)$ using group-specific weights, $P(G=g)$'s, that are obtained considering the relevance of each cohort over the total sample.
\\

Table \ref{tab:CS baseline} shows the baseline results on a dashboard of firm-level financial accounts\footnote{Please note that the number of observations includes, besides treated firms, all those firms for which we can observe markups and other relevant financial accounts. Therefore, the method applies inverse probability weights to develop a suitable control group. In the following tables, whenever we apply our baseline methodology, we will always report the total number of observations, as this is in the results of the procedure while indicating the number of treated firms in the table notes.}, including markups and other variables that can further qualify the impact of a takeover. After the treatment, most importantly, we do not observe a statistically significant impact on either markups or market shares. Apparently, in the latter case, after controlling for reverse causality, the statistically significant difference that we found in Table \ref{tab:descriptives} completely fades away. Evidently, we expected that previous positive associations were at least partially due to the screening capacity of acquirers, who could spot the targets with the higher economic potential. Yet, the lack of any significance on market power is counterintuitive, as one would expect that one of the main reasons why takeovers occur is to strengthen market position.

Still, we observe an increase in the scale of operations, evidenced by higher sales of about 2.1\% and variable costs of 2.6\%, as shown in columns (3) and (4), respectively. Notably, the variable cost ratio, as displayed in column (5), remains largely unchanged, implying that the rise in sales and variable costs is balanced, and both induce together an increase in operation volumes. In fact, we also observe that there is no significant impact on TFP and profitability. Instead, the dashboard suggests a marked decrease in the capital intensity of the targeted firm of about 5.1\% and remarkable financial stress derived from reduced liquidity (-7\%) and solvency ratios (8.7\%). The latter evidence could result from the acquiring company's strategic decisions to reduce unnecessary assets and use financial sources from the targeted subsidiary to pay back for the takeover deal. To investigate the baseline previous results further, we start by separating the sample for integration motives in the following paragraphs.

\begin{table}[htbp]
\caption {Average treatment effect on the treated (ATET) of takeovers on a dashboard of firm-level outcomes} \label{tab:CS baseline}
\begin{center}
\footnotesize
 \resizebox{\columnwidth}{!}{%
\begin{threeparttable}
\begin{tabular}{lcccccc} \hline
	&	(1)	&	(2)	&	(3)	&	(4)	&	(5)	\\
VARIABLES 	&	Markup	&	Market share	&	Sales	&	Variable cost	&	\makecell[c]{Variable cost \\ ratio} \\ \hline
ATET 	&	-0.002	&	0.013	&	0.021***	&	0.026***	&	0.005	\\
&	(0.002)	&	(0.008)	&	(0.006)	&	(0.007)	&	(0.003)	\\
Observations 	&	3,782,482	&	3,782,515	&	3,782,515	&	3,782,515	&	3,762,430	\\
Controls 	&	YES	&	YES	&	YES	&	YES	&	YES \\ \hline  
& (6)	&	(7)	& (8) & (9) & (10) \\ 
VARIABLES &	TFP	&	ROI & \makecell[c]{Capital \\ Intensity} & \makecell[c]{Liquidity \\ Ratio} & \makecell[c]{Solvency \\ Ratio} \\ \hline
ATET &	0.001	&	0.011	&	-0.051***	&	-0.070***	&	-0.087***	\\
&	(0.004)	&	(0.010)	&	(0.013)	&	(0.012)	&	(0.016)	\\
 Observations &	3,782,515	&	3,367,706	&	3,782,515	&	2,789,594	&	2,586,380	\\
Controls &	YES	&	YES	 & YES & YES & YES \\ 
\bottomrule
\end{tabular}
\begin{tablenotes}[para,flushleft]
\footnotesize{The table reports results following the difference-in-difference approach by \citet{Callaway_Santanna_2021}. ATET coefficients are obtained as a weighted average that considers the importance of each cohort of firms. The estimator is doubly robust, and the matching is obtained by inverse probability weighting controlling for firm size, age, capital intensity, TFP and 2-digit industry. There are 4,482 firms in the treated group, while the control group includes all the firms that have never been treated and those that have not been treated yet. Standard errors clustered at the firm level are in parentheses.  *** p$<$0.01, ** p$<$0.05, * p$<$0.1}.
\end{tablenotes}
\end{threeparttable}
}
\end{center}
\end{table}

\subsection{Vertical vs. horizontal integration strategies}\label{sec: integrations}

Our next step is to separate vertical and horizontal integration cases to identify whether heterogeneous markup changes stem from different integration strategies. The rationale is that there could be different mechanisms at play. Firms engaging in horizontal takeovers ab by sorb a direct competitor, possibly achieving a better market position. On the other hand, vertical integration strategies aim at absorbing a buyer or a supplier, possibly pursuing cost-saving strategies along a supply chain when intermediate inputs are delivered intra-firm after the acquisition. 

To spot horizontal mergers, we check whether the parent company and its subsidiary belong to the same industry at the 2-digit level of the NAICS 2002 classification. In the absence of data on direct shipments, to identify vertical integration, we follow \citet{fan_2000}, \citet{acemoglu_2009}, \citet{alfaro_2016}, and \citet{DelPrete_Rungi_2017}, who look at Input-Output coefficients derived from the Bureau of Economic Analysis (BEA). Thus, we compare the technical coefficients\footnote{Input-output technical coefficients, or input requirements, represent the amount of input from one sector required to produce one unit of output in another sector.} of the industry in which each subsidiary operates with the median coefficient of inputs required by the industry in which the parent company operates. We assume that a subsidiary and a parent company are in a vertical relationship if the I-O technical coefficients linking them through are above the sample median. Out of the 4,482 cases of acquisition in our sample, we distinguish 954 events of horizontal acquisitions and 1,955 of vertical acquisitions. Table \ref{tab:vert/hor} reports results for vertical and horizontal acquisitions\footnote{For vertical integration, horizontal acquisition cases are excluded from the sample and vice versa. Also, cases of alternative integration strategies, i.e. neither vertical nor horizontal, are excluded. Thus, in both cases, the control group includes companies that have never been acquired in our period of analysis. In addition, the control group includes companies that are not yet treated, i.e., they will be subject to vertical acquisitions in future periods for vertical cases and horizontal acquisitions for horizontal cases.} as shown in panels (a) and (b), respectively.

\begin{table}[htbp]
  \centering
  \caption{Average treatment effect on the treated (ATET) of takeovers: dashboards of vertical vs. horizontal integration strategies}\label{tab:vert/hor}
  \begin{subtable}[t]{\linewidth}
    \centering
    \footnotesize
    \vspace{0pt}
    \caption{Vertical integrations}
    \begin{tabular}{lccccc} \hline
		&	(1)	&	(2)	&	(3)	&	(4)	&	(5)	\\
VARIABLES 		&	Markup	&	Market share	&	Sales	&	Variable cost	&	\makecell[c]{Variable cost \\ ratio} \\ \hline
ATET 	&	-0.007**	&	0.025**	&	0.029***	&	0.032***	&	0.002	\\
&	(0.003)	&	(0.012)	&	(0.009)	&	(0.010)	&	(0.005)	\\
Observations &	3,749,600	&	3,749,633	&	3,749,633	&	3,749,633	&	3,729,558	\\
Controls 	&	YES	&	YES	&	YES	&	YES	&	YES \\ \hline  
& (6)	&	(7)	& (8) & (9) & (10) \\ 
VARIABLES &	TFP	&	ROI & \makecell[c]{Capital \\ Intensity} & \makecell[c]{Liquidity \\ Ratio} & \makecell[c]{Solvency \\ Ratio} \\ \hline
ATET &	0.003	&	0.023*	&	-0.072***	&	-0.108***	&	-0.110***	\\
&	(0.008)	&	(0.012)	&	(0.018)	&	(0.025)	&	(0.024)	\\
 Observations &	3,749,633	&	3,337,305	&	3,749,633	&	2,766,975	&	2,564,319	\\
Controls &	YES	&	YES	 & YES & YES & YES \\ 
\bottomrule
\end{tabular}
  \end{subtable}
  \begin{subtable}[t]{\linewidth}
    \centering
    \footnotesize
    \bigskip
    \caption{Horizontal integrations}
    \begin{threeparttable}
    \footnotesize
    \begin{tabular}{lccccc} \hline
    	&	(1)	&	(2)	&	(3)	&	(4)	&	(5)	\\
VARIABLES 	&	Markup	&	Market share	&	Sales	&	Variable cost	&	\makecell[c]{Variable cost \\ ratio} \\ \hline
ATET 	&	0.001	&	0.020	&	0.026*	&	0.034**	&	0.006	\\
&	(0.005)	&	(0.019)	&	(0.014)	&	(0.016)	&	(0.008)	\\
Observations 	&	3,736,372	&	3,736,405	&	3,736,405	&	3,736,405	&	3,716,336	\\
Controls 	&	YES	&	YES	&	YES	&	YES	&	YES \\ \hline  
        & (6) &	(7)	& (8) & (9) & (10) \\ 
VARIABLES &	TFP	&	ROI & \makecell[c]{Capital \\ Intensity} & \makecell[c]{Liquidity \\ Ratio} & \makecell[c]{Solvency \\ Ratio} \\ \hline
ATET &	0.007	&	0.010	&	-0.028	&	-0.121***	&	0.014	\\
&	(0.008)	&	(0.027)	&	(0.029)	&	(0.023)	&	(0.032)	\\
 Observations &	3,736,405	&	3,325,084	&	3,736,405	&	2,757,150	&	2,554,679	\\
Controls &	YES	&	YES	 & YES & YES & YES \\ 
\bottomrule
    \end{tabular}
    \begin{tablenotes}[para,flushleft]
\footnotesize{The table reports results following the difference-in-difference approach by \citet{Callaway_Santanna_2021}. ATET coefficients are obtained as a weighted average that considers the importance of each cohort of firms. The estimator is doubly robust, and the matching is obtained by controlling for firm size, age, capital intensity, TFP and 2-digit industry. There are 1,955 firms in the treated group for vertical integrations and 954 treated firms for horizontal integrations. The control group includes firms that have never been treated and those that have not been treated yet. Standard errors clustered at the firm level are in parentheses.  *** p$<$0.01, ** p$<$0.05, * p$<$0.1}.
\end{tablenotes}
\end{threeparttable}
  \end{subtable}
\end{table}


By looking at horizontal takeovers in panel (b) of Table \ref{tab:vert/hor}, once again, we find that takeovers do not have any significant impact either on markups or market shares. This is similar to the general case of takeovers, before separating the motivation of the deals. We believe this an important result, as it could indicate that competition policy in the European Union successfully limits market abuses (hence, market power) after screening mergers and takeovers. Nonetheless, we find evidence that both sales and variable costs are higher after a horizontal integration, pointing to an overall impact on the volume of activity, hence firm size, but not on productivity and profitability, somehow in line with evidence of the baseline exercise in \ref{tab:vert/hor}.\\

Most importantly, we find a small albeit significant average impact of takeovers on markups in the case of vertical integration strategies. In particular, we observe in Table \ref{tab:vert/hor}, panel (a), that the acquired firm records on average a 0.7\% lower markup, together with an increase in sales of 2.9\%, in variable costs of 3.2\%, and a decrease in capital intensity of 7.2\%.\\

\textit{Prima facie}, the gains stemming from vertical integration can be directly related to the access of either tangible or intangible inputs at a lower cost \citep{atalay_2014}. They can also result from more efficient use of technology and logistics \citep{Hortaccsu_Syverson_2007}. Notably, in our case, we find robust evidence that targeted firms in the European Union increase their scale of operations after the takeover, but, differently from previous works \citep{McGukin_Nguyen_1995, Gugler_et_al_2003, Maksimovic_et_al_2011}, we do not find evidence of an impact on productivity, while capital intensity becomes lower, possibly as a result of industrial restructuring. 

Interestingly, we find that liquidity ratios show increasing financial constraints for the targeted firms in both vertical and horizontal integration strategies. The ratios we analyze are intended to assess the financial stability of the firm. The solvency ratio considers all of a company's assets and how they can cover liabilities, including long-term debt with a maturity of more than one year, while the liquidity ratio takes into account only the most liquid assets, such as cash and marketable securities, and how these can be used to cover upcoming obligations in the short term. Our hypothesis is that acquired targets often have to repay for at least a part of the takeover deals.

Overall, the acquiring firm’s strategy centers on expanding the target’s scale, lowering markups, and extracting value through vertical integration. By eliminating double marginalization, the firm reduces markups, and thereby reinforces cost efficiency. Meanwhile, rising market share and sales indicate effective coordination and resource sharing, reflecting successful exploitation of scale advantages. The decline in capital intensity, coupled with stable cost ratios, points to reconfigured production processes that economize on physical assets. However, the simultaneous worsening of liquidity and solvency suggests a shift of financial risk towards the target, highlighting the acquirer's emphasis on operational synergies and financial burden relief.

Finally, we unroll the impact of takeovers throughout our period of analysis\footnote{Please note that we report only the variables from the dashboard of financial accounts that we believe are relevant for our discussion on the unrolling of market power. Further details are available upon request.}. Therefore, we perform an event study analysis, for which we need to compute the length of exposure to the treatment. In plain words, our event study returns the average impact on subsidiaries' outcomes after $e$ periods from being taken over by a parent company. Following \citet{Callaway_Santanna_2021}, it takes into account the heterogeneity of different cohorts of takeovers that participate in the treatment group. Visual evidence of the results for vertical takeovers is shown in Figure \ref{fig:event}, while we provide evidence of parallel trends test in Table \ref{tab:parallel_trend} of Appendix B. Notably, we record that markups tend to decrease more over time, in the years following the takeover, up to almost 4\% by year in the case of that group of subsidiaries that we observe for the longest period.

\begin{figure}[H]
\caption{Event study after vertical takeovers}
\begin{minipage}{\textwidth}
\begin{center}
\includegraphics[scale=0.14]{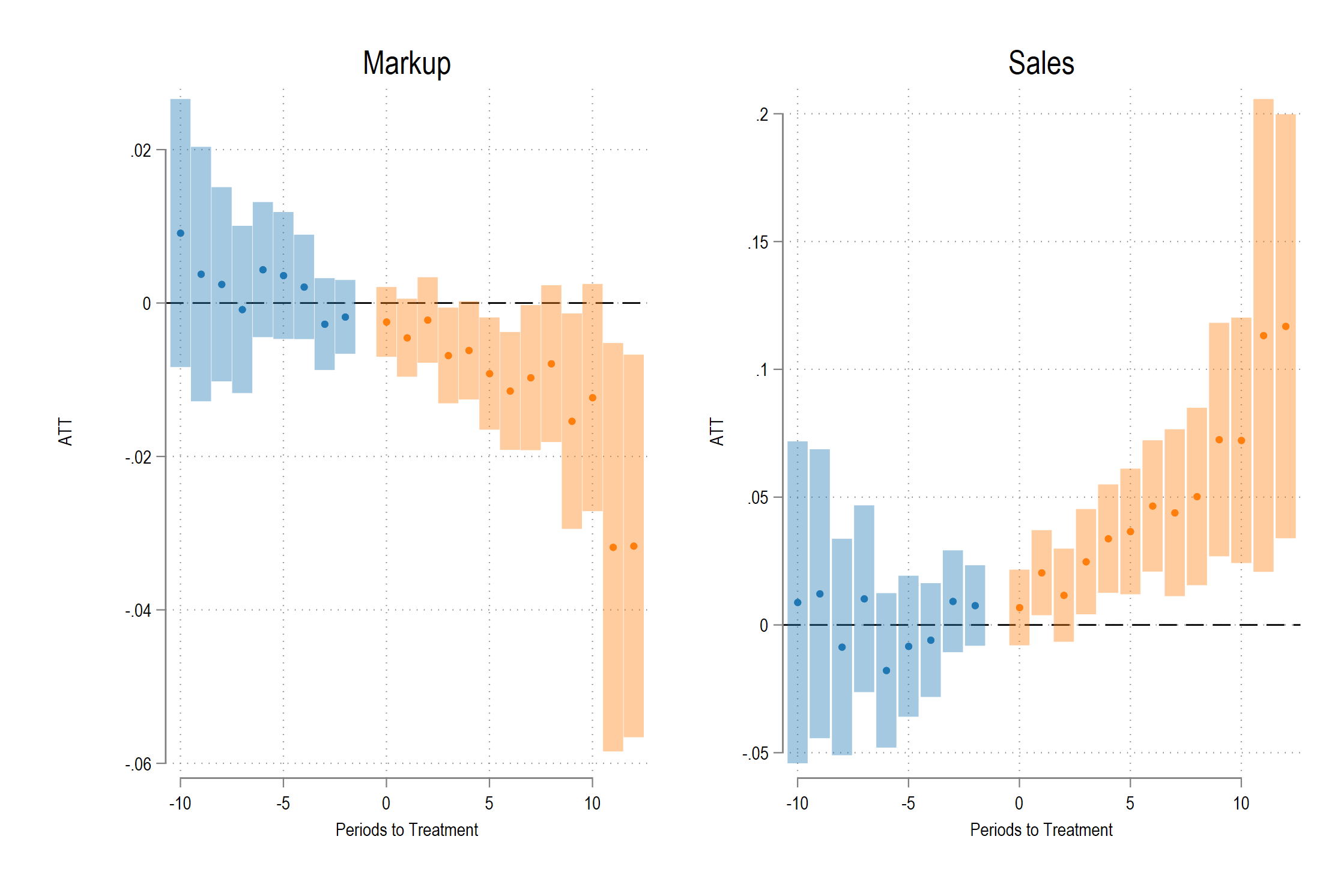}
\includegraphics[scale=0.14]{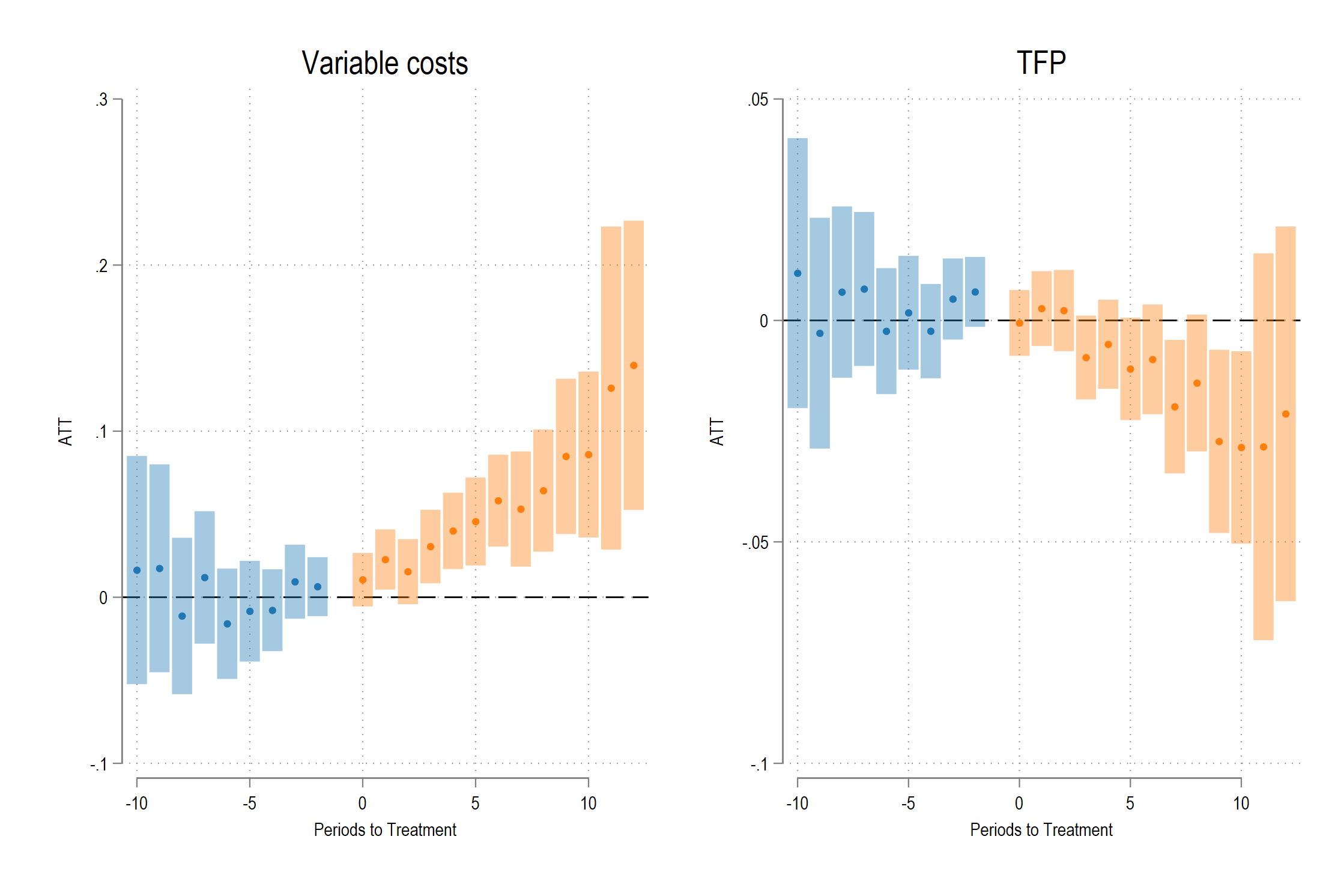}
\includegraphics[scale=0.14]{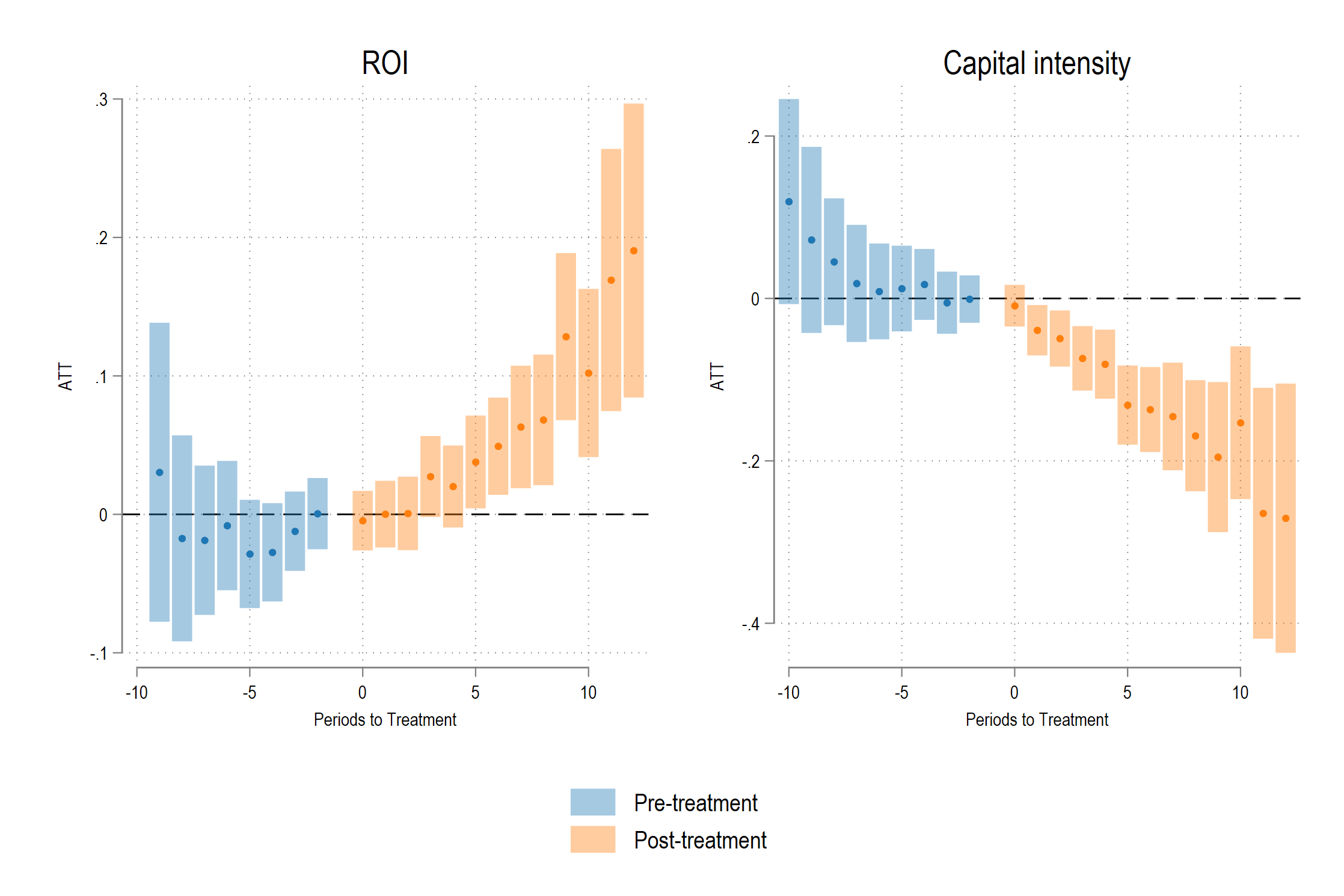}
\end{center}
\begin{tablenotes}[para,flushleft]
\footnotesize{The figure shows the effect of vertical takeovers in an event study setting following \citet{Callaway_Santanna_2021}. The event study plots consider symmetric differences before and after the treatment.}
\end{tablenotes}
\end{minipage}
\label{fig:event}
\end{figure}

\subsection{On the elimination of double margins after vertical integration}

The simultaneous occurrence of decreasing markups, increasing market shares, and increasing volumes of activity (Table \ref{tab:vert/hor} and Figure \ref{fig:event}) suggest that the elimination of double profit margins is the main mechanism at work after vertical takeovers. The textbook case is a setting in which an upstream supplier has some monopolistic power, and she confronts a downstream monopolistic buyer along the supply chain. In this case, when we have independent firms, the
downstream buyer adds her markup on the product whose inputs were already
marked up by the upstream supplier. That is, double (or multiple) profit margins sum up along the supply chain, making the final goods more expensive for consumers. Then, if vertical integration occurs between the two firms, they can eliminate double marginalization, thus lowering prices and increasing sales. In this setup, both the producers and the consumers benefit from the vertical takeover thanks to the elimination of deadweight losses.

Previous works have already pointed to the existence of pro-competitive effects as a consequence of vertical integration strategies \citep{chipty_2001,Gil_2015, crawford_2018}. In this context, the presence of a double profit marginalization before vertical integration is seen as a market externality whose persistence along supply chains has a negative impact on social welfare. 

Yet, despite the welfare efficiency gains stemming from the elimination of double margins, we cannot exclude those vertical integrations might still create distortions through the foreclosure of other competitors or a strategic rise in prices of other goods or services in a portfolio of multiproduct firms \citep{Spengler_1950, luco_2020}. It is beyond the scope of our analyses to investigate whether there is indeed a general equilibrium welfare effect in Europe from takeover activities. Yet, we refer to the theoretical work by \citet{chone_2023}, who discuss how eliminating double profit margins can have side effects. Depending on the distribution of the bargaining power among the parts involved in the acquisition, a vertical integration strategy can threaten the market position of the other independent suppliers, therefore leading to foreclosure effects. When the buyer has full bargaining power over prices and quantities, the vertical acquisition always benefits final consumers, while in cases of reduced bargaining power, after the buyer has committed to deal exclusively with a more limited set of suppliers, the exclusion of efficient suppliers potentially harms final consumers. 

Notably, competition policy analysts have been discussing specific case studies of double marginalization that have been brought to the attention of competition authorities. And yet, recent years have seen a revamped debate on the approach to take against vertical takeovers \citep{Dewulf_et_al_2022}, especially after the publication of two different updates of the U.S. Vertical Merger Guidelines in 2020 and 2023. The arguments are the same as those we just sketched in the above paragraphs. On one side, we have the advocates for procompetitive effects thanks to the elimination of deadweight losses. On the other side, we have the scepticism of analysts who underline the harm that vertical integration can still cause once considering indirect effects on the market structure.

\subsection{Parent companies and heterogeneous supply chains}

Up to now, we have focused exclusively on the impact on targeted firms, and we have neglected what happens to acquirers. Please note that the textbook case of double marginalization includes both. In the following paragraphs, we will first check changes in the dashboard of firm-level accounts of those parent companies that completed the takeovers in our period of analyses. Then, we will check what happens when we introduce heterogeneous supply chains by parent companies, i.e., when the extent of the subsidiaries that are already integrated varies before the new takeover occurs. Our intuition is that the newly acquired company can potentially trade intermediate inputs with the entire corporate perimeter once the takeover is complete.

First, as a basic test, we define as treated a parent\footnote{Please note, however, that it is more appropriate to define the treatment with respect to what happens to subsidiaries, as the latter are the ones that eventually change ownership and, with it, they change their approach to the market. Parent companies can integrate multiple subsidiaries along the timeline, and we cannot expunge all the noise that multiple vertical integrations can bring to their consolidated financial data.} that has acquired a majority equity stake in at least one subsidiary in our period of analysis. Consequently, we will match a control group made up of parent companies that did not take over (or they did not take over yet) any company during our analysis period. After following our baseline empirical setup described in Section \ref{sec: strategy}, we report results in Table \ref{tab:shareholders} for the entire sample, and separately for vertical integration strategies in Table \ref{tab:shareholders_vertical}. 

Notably, we do not find any significant change in either of the financial accounts in the general case of Table \ref{tab:shareholders}, with the exception of a weakly negative significant coefficient in profitability (ROI: 5\%). On the other hand, we find some positive effects on market shares and sales but no impact on markups. From our point of view, the (lack of) findings on parent companies are compatible with eliminating double margins, although they do not fit the textbook case. In fact, according to textbook theory, the reduction of markups could occur on both sides of the deal, i.e., the buyer and the supplier of intermediate inputs. The main difference between the standard case and what we have in our data is that in the first case, there is an implicit assumption that neither the buyer nor the supplier engaged in any previous vertical integration. Yet, in reality, when we look at sample parents, they often already coordinate dozens or hundreds of subsidiaries before they engage in a new takeover. See preliminary evidence in Figure \ref{fig: corporate_perimeter}. Please note that, according to our data, the average number of subsidiaries already integrated by sample parents is eighty-eight, while the median is six. 

On the one hand, parent companies with already integrated subsidiaries may benefit from a higher bargaining power and, thus, lead newly acquired subsidiaries to unilaterally reduce their markups. On the other hand, from a simple accounting perspective, bigger parent companies can consolidate the financial activities of old and new subsidiaries, and thus the presence of non-significant coefficients in Tables \ref{tab:shareholders} and \ref{tab:shareholders_vertical} are just the consequence of composition effects at the level of subsidiaries.

\begin{table}[htbp]
\caption {Average treatment effect on the treated (ATET) of takeover: a dashboard of financial accounts of parent companies} \label{tab:shareholders}
\begin{center}
\footnotesize
 \resizebox{\columnwidth}{!}{%
\begin{threeparttable}
\begin{tabular}{lccccc} \hline
		&	(1)	&	(2)	&	(3)	&	(4)	&	(5)	\\
VARIABLES 	&	Markup	&	Market share	&	Sales	&	Variable cost	&	\makecell[c]{Variable cost \\ ratio} \\ \hline
ATET 	&	0.006 & -0.031 & 0.006 & 0.012 & 0.004 \\
 	&	(0.010)	& (0.059) & (0.024) & (0.027) & (0.010)	\\
Observations 	&	52,597	& 53,536 & 53,536 & 53,536 & 53,331 \\
Controls 	&	YES	&	YES	&	YES	&	YES	&	YES \\ \hline  
& (6)	&	(7)	& (8) & (9) & (10) \\ 
VARIABLES &	TFP	&	ROI & \makecell[c]{Capital \\ Intensity} & \makecell[c]{Liquidity \\ Ratio} & \makecell[c]{Solvency \\ Ratio} \\ \hline
ATET & 0.003 & -0.050*  & 0.001 & -0.031 & 0.010 \\
& (0.014) & (0.026)	&  (0.042) & (0.027) & (0.025) \\
 Observations &	53,536 & 47,191 & 53,536 & 53,242
 & 52,523 \\
Controls &	YES	& YES & YES & YES & YES \\ 
\bottomrule
\end{tabular}
\begin{tablenotes}[para,flushleft]
\footnotesize{The table reports aggregate results obtained following the methodological approach proposed by \citet{Callaway_Santanna_2021} to account for heterogeneity in treatment timing. Single coefficients of the ATET are obtained with a weighted average that considers the importance of each cohort of firms at different times. Estimations are obtained through a doubly robust estimator and include firms' characteristics as control variables. There are 3,776 treated parent companies, and the control group is composed of never-treated units and not-yet-treated units. Variables are in logs. Standard errors clustered at the firm level are reported in parentheses, and significance levels are *** p$<$0.01 ** p$<$0.05 * p$<$0.1}
\end{tablenotes}
\end{threeparttable}
}
\end{center}
\end{table}

\begin{table}[htbp]
\caption {Average treatment effect on the treated (ATET) of vertical takeovers: a dashboard of financial accounts of parent companies} \label{tab:shareholders_vertical}
\begin{center}
\footnotesize
 \resizebox{\columnwidth}{!}{%
\begin{threeparttable}
\begin{tabular}{lccccc} \hline
		&	(1)	&	(2)	&	(3)	&	(4)	&	(5)	\\
VARIABLES 	&	Markup	&	Market share	&	Sales	&	Variable cost	&	\makecell[c]{Variable cost \\ ratio} \\ \hline
ATET 	&	-0.006	&	0.053***	&	0.141*	&	0.139	&	-0.0003	\\
&	(0.028)	&	(0.165)	&	(0.082)	&	(0.096)	&	(0.028)	\\
Observations 	&	42,393	&	43,147	&	43,147	&	43,147	&	42,962	\\
Controls 	&	YES	&	YES	&	YES	&	YES	&	YES \\ \hline  
& (6)	&	(7)	& (8) & (9) & (10) \\ 
VARIABLES &	TFP	&	ROI & \makecell[c]{Capital \\ Intensity} & \makecell[c]{Liquidity \\ Ratio} & \makecell[c]{Solvency \\ Ratio} \\ \hline
ATET &	-0.047	&	46.445	&	-0.038	&	0.077	&	-0.018	\\
&	(0.047)	&	(28.441)	&	(0.126)	&	(0.133)	&	(0.095)	\\
 Observations &	43,147	&	37,994	&	43,147	&	42,923	&	42,298	\\
Controls &	YES	& YES & YES & YES & YES \\ 
\bottomrule
\end{tabular}
\begin{tablenotes}[para,flushleft]
\footnotesize{The table reports aggregate results obtained following the methodological approach proposed by \citet{Callaway_Santanna_2021} to account for heterogeneity in treatment timing. Single coefficients of the ATET are obtained with a weighted average that considers the importance of each cohort of firms at different times. Estimations are obtained through a doubly robust estimator and include firms' characteristics as control variables. There are 2,305 treated parent companies, and the control group is composed of never-treated units and not-yet-treated units. Variables are in logs. Standard errors clustered at the firm level are reported in parentheses, and significance levels are *** p$<$0.01 ** p$<$0.05 * p$<$0.1}
\end{tablenotes}
\end{threeparttable}
}
\end{center}
\end{table}

Eventually, it is more relevant to us to test whether the heterogeneity of the supply chain already integrated by a parent company has any role. We have observed in Figure \ref{fig: corporate_perimeter} that a few parent companies may already control hundreds of subsidiaries before integrating the ones in our sample. Interestingly, we have already shown a negative association between the number of previously integrated subsidiaries and the impact on new subsidiaries' markups in Figure \ref{fig: association markups and the corporate perimeter}. Here, we want to go beyond the correlations and check whether such a negative association stands challenging causality. 

\begin{figure}[H]
\caption{Impact on markups of targeted subsidiaries in relationship with the number of subsidiaries already integrated by acquiring parent companies}
\centerline{\includegraphics[scale=0.8]{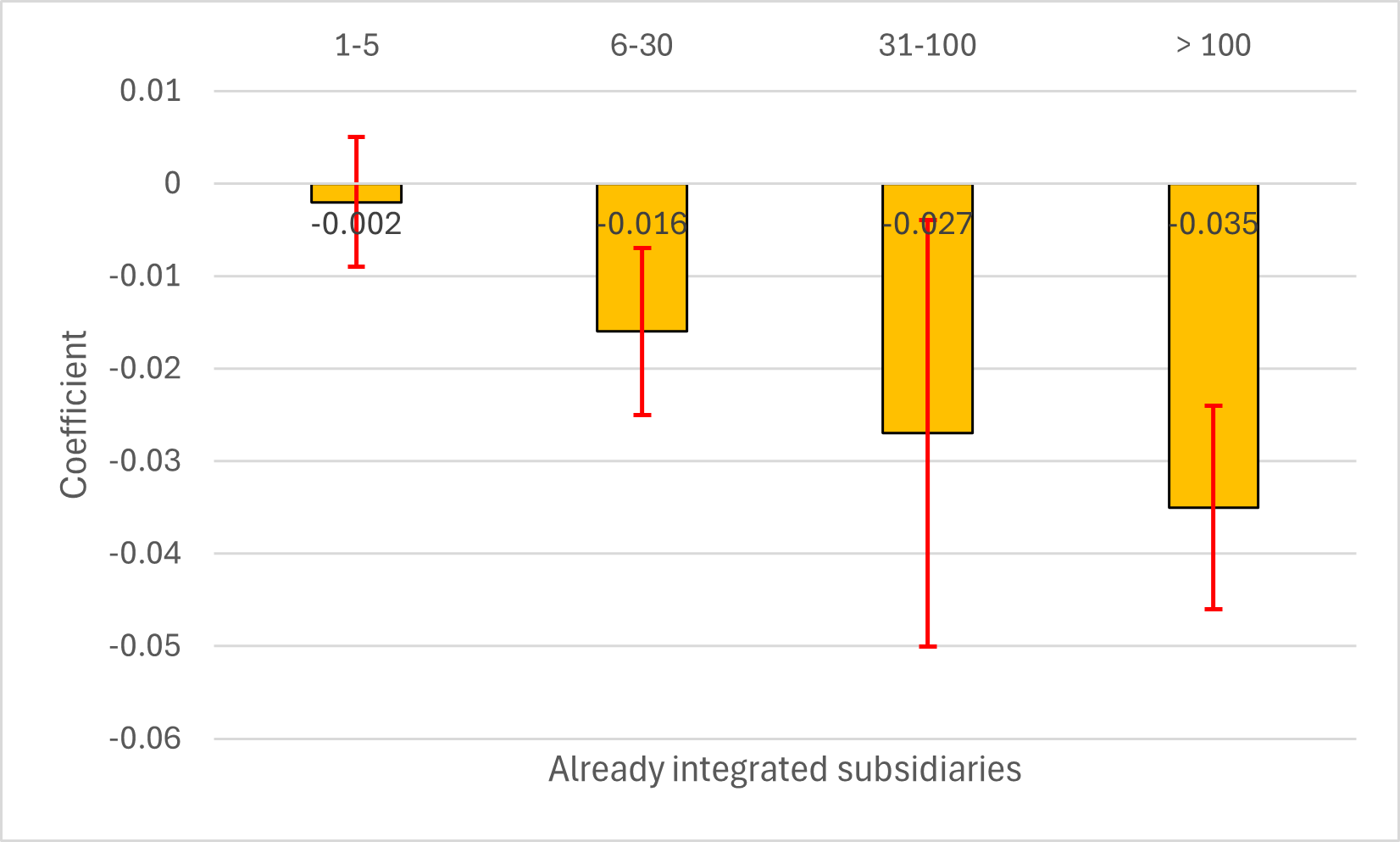}}
 \begin{tablenotes}
      \footnotesize
      \singlespacing
      \item Note: The figure shows the results of four different implementations of our baseline exercise based on the experience by \citet{Callaway_Santanna_2021} with pre-treatment trends after we separate cases of up to 5, from 6 to 30, from 31 to 100, and more than 100 co-affiliates in the same corporate perimeter. Bars indicate the Average Treatment Effect on the Treated (ATET) for each category, calculated considering a panel set up in 2007-2021. The control group is matched with inverse probability weights (IPW) starting from companies that were never controlled in our study by any parent company. Bars indicate 95\% confidence intervals.
   \end{tablenotes}
\label{fig: markups and corporate_perimeter}
\end{figure}

\begin{figure}[H]
\caption{Impact on sales of targeted subsidiaries in relationship with the number of subsidiaries already integrated by acquiring parent companies}
\centerline{\includegraphics[scale=0.8]{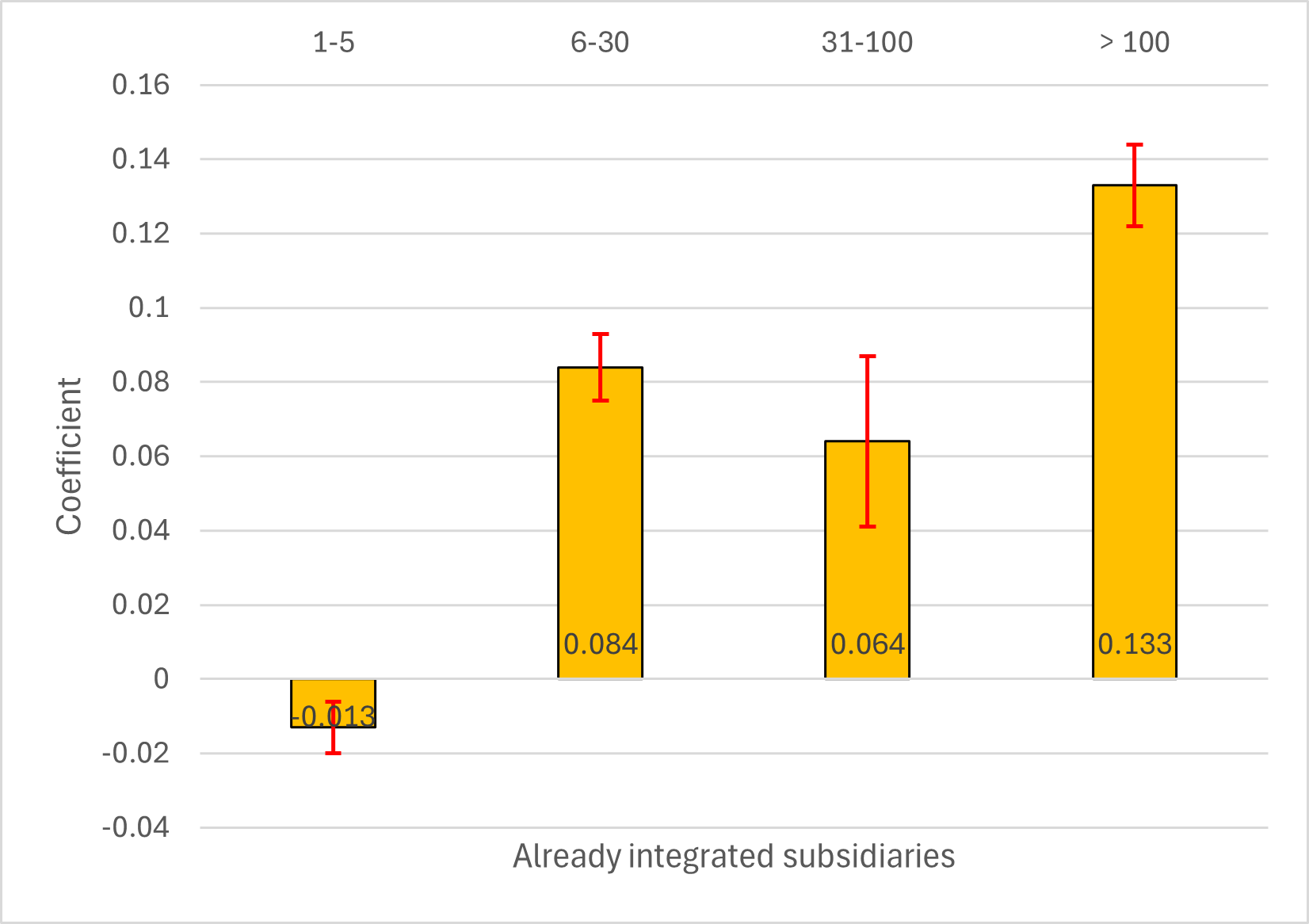}}
 \begin{tablenotes}
      \footnotesize
      \singlespacing
      \item Note: The figure shows the results of four different implementations of our baseline exercise based on the experience by \citet{Callaway_Santanna_2021} with pre-treatment trends after we separate cases of up to 5, from 6 to 30, from 31 to 100, and more than 100 co-affiliates in the same corporate perimeter. Bars indicate the Average Treatment Effect on the Treated (ATET) for each category, calculated considering a panel set up in 2007-2021. The control group is matched with inverse probability weights (IPW) starting from companies that were never controlled in our study by any parent company. Bars indicate 95\% confidence intervals.
   \end{tablenotes}
\label{fig: sales and corporate_perimeter}
\end{figure}

In this case, we adopt our baseline specification in eq. \ref{eq: aggregate_ATT} to four subsamples, separating treated subsidiaries if they belong to parent companies with 1-5, 6-30, 31-100, and higher than 100 subsidiaries, respectively. What we find is that previous findings on decreasing markups are mainly driven by subsidiaries that, after the deal, belong to parent companies with at least six already integrated subsidiaries. Results on markups are visualized in Figure \ref{fig: markups and corporate_perimeter}, and those on sales are visualized in Figure \ref{fig: sales and corporate_perimeter}. If we consider the number of subsidiaries as a proxy for the extent of integrated supply chains, we conclude that there is a tendency for markups to be lower when they integrate into bigger corporate perimeters. This is consistent with the idea that the elimination of double margins is at work, and it works especially more when the chances to source inputs from within the corporate perimeter by the acquirer are higher. Yet, the elimination of double profit margins allows companies to improve their market position, and that's why we also observe increasing sales that are all the more increasing when the acquiring parent has already integrated a bigger segment of supply chains, as shown in bars in Figure \ref{fig: sales and corporate_perimeter}. Positive and significant impacts are recorded when there are at least six co-affiliates (between 6.4\% and 8.4\%), yet, in the case of $>100$ co-affiliates, the impact on sales is higher, on average about 13.3\%.

\section{Robustness and sensitivity analysis}\label{sec: robustness}

In the following paragraphs, we perform a battery of robustness and sensitivity checks starting from the previous baseline analyses. For convenience, we separate into checks stemming from methodological concerns, on the one hand, and checks that deal with possible sample composition effects, on the other hand. 

\subsection{Methodological concerns}
The first concern relates to the methodology we adopt to estimate markups. We rely on the production function approach following \citet{DeLoecker_Warzynski_2012}, who adapt the cost-based approach initially developed by \citet{Hall_1988}. For more details, see Appendix A. The advantage of this method is that it requires minimal data and relatively weak assumptions applied to wider datasets of firm-level financial accounts. Nevertheless, pitfalls have been discussed by \citet{Basu_2019}, \citet{syverson_2019} and \citet{traina_2018}. Recent work by \citet{bond_2021} highlights that identification and estimation issues persist when firm-level output prices are not directly observed. For this reason, one may prefer industry-specific \textit{ad-hoc} subsets, including precise quantities and prices. Information that is not available in common firm-level financial accounts. In previous results, we partially address the omitted price bias in two ways. On the one hand, we convert monetary values to quantities after using industry-specific price deflators. On the other hand, we estimate the output elasticity of materials across sectors by holding a fixed time dimension, i.e., assuming that materials are a flexible input and that there are no adjustment costs. In this way, as already pointed out by \citet{DeLoecker_2021}, the change in the ratio of revenue to the materials' expenditure is a direct estimate of the change in the markup. 
 
Yet, one important element in the estimation of markups is the choice of what the flexible inputs are. We perform a sensitivity analysis, for which we show what happens when we modify our choice. In fact, in the previous sections, we considered only intermediate inputs (materials), following previous suggestions from economic literature, according to which labour inputs are less flexible than materials in institutional contexts different from the United States. When a representative producer in the European Union has to choose a combination of factors of production, she will probably encounter more friction than her colleagues in the United States. Therefore, we present new results in Table \ref{table:rob_markups} assuming that: i) only the labour inputs are flexible (columns (1) and (3)); ii) both intermediate inputs and labour costs participate in a composite variable input (columns (2) and (4)). The latter is constructed as the simple sum of the cost of materials and the cost of employees as in \citet{raval_2023}\footnote{\citet{raval_2023} finds that firm-level markups estimated using labour and materials, alternatively, are negatively correlated and have opposite time trends, possibly due to non-neutral productivity differences across firms.}. In this case, as expected, the markup changes are either smaller than the baseline when we look at the composite input, or they are not significant if we consider only labour as the flexible input. As per previous discussions, we prefer to keep our baseline estimates where intermediate inputs (materials) are considered flexible inputs.
 
\begin{table}[h]
\centering
\caption{Average treatment effect on the treated (ATET) on markups after vertical takeovers: sensitivity to measures of markups}\label{table:rob_markups}
\footnotesize
\begin{threeparttable}
\begin{tabular}{lcccc}
\hline
& \multicolumn{2}{c}{Baseline} & \multicolumn{2}{c}{Vertical} \\
\cmidrule(l){2-3} \cmidrule(l){4-5}
& (1) & (2) & (3) & (4) \\
 & Markup Labor & \makecell[c]{Markup \\ composite input} & Markup Labor & \makecell[c]{Markup \\ composite input} \\
\hline
ATET & -0.003 & -0.001** & -0.001 & -0.002** \\
& (0.006) & (0.000) & (0.009) & (0.001) \\
Observations & 3,759,377 & 3,782,515 & 3,727,122
 & 3,749,633 \\
\bottomrule
\end{tabular}
\begin{tablenotes}[para,flushleft]
\footnotesize{Note. The table shows results using estimations of firm-level markups \'a la \citet{DeLoecker_Warzynski_2012} relying on different variable inputs: columns (1) and (3) report results using labour costs for the baseline and vertical integrations, respectively. The estimates are obtained after a diff-in-diff with a panel set up following \citet{Callaway_Santanna_2021}. Columns (2) and (4) report results using a composite input from the sum of labour and material costs. Variables are in logs. There are 1,955 treated firms, while the control group includes firms that have never been treated and those that have not been treated yet. Standard errors clustered at the firm level are reported in parentheses, and significance levels are *** p$<$0.01 ** p$<$0.05 * p$<$0.1}
\end{tablenotes}
\end{threeparttable} 
\end{table}

A second concern is once again methodological. Markups are derived following a production function approach as in \citet{Callaway_Santanna_2021} after we estimate production functions using the methodology proposed by \citet{Ackerberg_Caves_Frazer_2015}, which handles the endogeneity of productivity estimates when adjustment to shocks on inputs markets are simultaneous. As an additional robustness check, we compute our findings with the output elasticity obtained after a simple OLS estimation of the revenue production function. Column (1) of Appendix Table \ref{tab:ols} shows the results. We get consistent coefficients as subsidiary firms subject to vertical takeovers still reduce their level of markups by about 0.5\%.\\

A third methodological concern is the adoption of a panel setting for our identification strategy. Our preferred approach \'a la \citet{Callaway_Santanna_2021} is able to catch variation in treatment timing, as we explained in Section \ref{sec: strategy}. Yet, we may want to compare with a more classical combination of propensity score matching with a two-period difference-in-difference. For our purpose, we first derive a control group made of firms with similar characteristics, which we use as a counterfactual for the absence of treatment. Our aim is to control for potential self-selection of firms into a treatment status, as in the case of cherry-picking by parent companies that screen for targets with the best economic potential. We implement our propensity score matching using a one-nearest neighbour matching scheme with the assumption of a common support\footnote{Unlike the matching process implemented in the main analysis, in which all observations are retained and have an assigned weight, in this case, we use a subsample in which for each treated unit, one control unit is associated (nearest-neighbour) thanks to the similarity of the p-scores.}. The match is obtained after a logit regression that predicts the treatment's probability based on firms' size, capital intensity, productivity, age, country of origin, and affiliation to a 2-digit industry in a given year. The exercise is repeated separately for horizontal and vertical takeovers, respectively. To assess the performance of the matching procedure, Appendix Figure \ref{fig:balancing properties} shows the balancing properties between the treatment and the control groups, including standardized \% bias and the variance ratios across covariates before and after matching the entire sample. See \citet{RosenbaumRubin1}, \citet{RosenbaumRubin2}, and \citet{Rubin} for further details. Once we have a suitable control group, we proceed by estimating the usual difference-in-difference specification on our matched sample:

\begin{equation}
y_{i,t}= \beta_0 + \beta_1 T_{i} + \beta_2 Post_{i,t} + \beta_3 T_{i}*Post_{i,t} + \beta_4X_{i,t} + \gamma_t + \delta_k + \omega_l + \epsilon_{i,t}
\end{equation}

where $y_{i,t}$ represents the logarithm of the outcome variables (markups, in our case), $T_{i}$ is a dummy to identify treated firms, $Post_{i,t}$ is a dummy variable equal to 1 if the firm has been the target of a takeover at time $t$. In the above specification, $\gamma_t$, $\delta_k$ and $\omega_l$ represent fixed effects for years, countries, and 2-digit NACE rev. 2 sectors, respectively, while $X_{i,t}$ is a set of control variables including capital intensity, age, TFP and firm size. $\beta_3$ is our coefficient of interest, indicating the effect of the takeover on the outcome variable capturing the average difference between treated firms before and after the treatment. Results after the propensity score matching are reported in columns (2) and (4) of Table \ref{tab:psm}. We find in columns (2) and (4) that both horizontal and vertical takeovers bring about a decrease in average markups with magnitudes 4.3\% and 3.8\%, respectively, which are higher than in our baseline panel setup.  In columns (1) and (3), we also include raw results before the matching, i.e., when non-treated observations are not filtered, and yet we include covariates in the specification (capital intensity, TFP, firm age, firm size). Coefficients are all negative with a higher magnitude. Please note, however, that we prefer to keep our baseline findings with a panel set-up, as introduced in the previous sections because we acknowledge distortions if we do not take care of the multiple treatment timing.

\begin{table}[htbp]
\caption {Average treatment effects on the treated (ATET) markups: sensitivity to methodologies} \label{tab:psm} \begin{center}
\footnotesize
\begin{threeparttable}
\begin{tabular}{lcccccc} \hline
& \multicolumn{2}{c}{Horizontal} & \multicolumn{2}{c}{Vertical} \\
\cmidrule(l){2-3} \cmidrule(l){4-5}
 & (1)  & (2) & (3) & (4) \\
VARIABLES & Markup (I) &  Markup (II) & Markup (III) & Markup (IV) \\ \hline
ATET & -0.057*** & -0.043*** & -0.065*** & -0.038***  \\
 & (0.002) & (0.003) & (0.002) & (0.002) \\ \hline
Obs. & 46,155 & 46,155 & 28,603 & 28,603 \\
Untreated obs. &  3,518,681 &  3,518,681 & 3,690,857 & 3,690,857\\
Treated obs. & 12,348 & 12,348 & 25,576 & 25,576\\
Matching & No & Yes & No & Yes\\
Controls & Yes & Yes & Yes & Yes \\
\hline
\end{tabular}
\begin{tablenotes}[para,flushleft]
\footnotesize{The table shows results on markups adopting a two-way diff-in-diff in columns (1) and (3) and after a propensity score matching in columns (2) and (4). Matching is implemented for nearest neighbors with logit specifications, including controls for the endogenous selection into takeovers: capital intensity, firm age, firm size, TFP, NACE 2-digit industries, and origin countries. Standard errors are reported in parentheses, and significance levels are *** p$<$0.01 ** p$<$0.05 * p$<$0.1}
\end{tablenotes}
\end{threeparttable}
\end{center}
\end{table}

Lastly, we are aware that our results could be sensitive to the Input-Output Table's threshold used to define vertical integration. To test the robustness of our findings, we propose to check what happens when we vary the threshold, adopting either a 25\% cutoff to capture a broader definition of vertical integration, or a 75\% cutoff to isolate cases where firms are strongly linked through technological complementarities. The results, reported in Table \ref{tab:vert_pctile} align with expectations. When we apply the stricter 75\% threshold, the negative effect on markups becomes even stronger than baseline results. By contrast, when we lower the threshold to 25\%, thereby including weaker forms of integration, we find no significant effect on markups, suggesting that a limited scope for efficiency gains. Additional findings — such as the increase in market shares, the proportional rise in sales and variable costs, as well as the reduction in capital intensity and the higher financial distress — remain consistent with the baseline estimates. We believe that such consistency suggests that effects are primarily driven by the acquisition itself, rather than by the extent of vertical integration achieved.

\begin{table}[H]
  \centering
  \caption{Average treatment effect on the treated (ATET) of takeovers at different thresholds of vertical integration}
  \label{tab:vert_pctile}
  
  \begin{subtable}[t]{\linewidth}
    \centering
    \footnotesize
    \caption{Vertical integrations – 25th percentile}
    \begin{tabular}{lccccc} \toprule
    & (1) & (2) & (3) & (4) & (5) \\
    VARIABLES & Markup & Market Share & Sales & Variable Cost & \makecell[c]{Variable cost \\ ratio} \\ \midrule
    ATET      & -0.004 & 0.018* & 0.022*** & 0.025*** & 0.004 \\
              & (0.003) & (0.011) & (0.008) & (0.009) & (0.004) \\
    Observations & 3,756,881 & 3,756,914 & 3,756,914 & 3,756,914 & 3,736,836 \\
    Controls  & YES & YES & YES & YES & YES \\ \midrule
    & (6) & (7) & (8) & (9) & (10) \\
    VARIABLES & TFP & ROI & \makecell[c]{Capital \\ Intensity} & \makecell[c]{Liquidity \\ Ratio} & \makecell[c]{Solvency \\ Ratio} \\ \midrule
    ATET      & 0.003 & 0.023** & -0.063*** & -0.013 & -0.121*** \\
              & (0.006) & (0.011) & (0.016) & (0.035) & (0.030) \\
    Observations & 3,756,914 & 3,344,013 & 3,756,914 & 2,771,960 & 2,569,217 \\
    Controls  & YES & YES & YES & YES & YES \\
    \bottomrule
    \end{tabular}
  \end{subtable}
  
  \begin{subtable}[t]{\linewidth}
  \centering
  \begin{threeparttable}
    \footnotesize
    \caption{Vertical integrations – 75th percentile}
    \begin{tabular}{lccccc} \toprule
    & (1) & (2) & (3) & (4) & (5) \\
    VARIABLES & Markup & Market Share & Sales & Variable Cost & \makecell[c]{Variable cost \\ ratio} \\ \midrule
    ATET      & -0.013*** & 0.037** & 0.038*** & 0.040*** & 0.005 \\
              & (0.004)   & (0.017) & (0.013)  & (0.014)  & (0.006) \\
    Observations & 3,736,473 & 3,736,506 & 3,736,506 & 3,736,506 & 3,716,438 \\
    Controls  & YES & YES & YES & YES & YES \\ \midrule
    & (6) & (7) & (8) & (9) & (10) \\
    VARIABLES & TFP & ROI & \makecell[c]{Capital \\ Intensity} & \makecell[c]{Liquidity \\ Ratio} & \makecell[c]{Solvency \\ Ratio} \\ \midrule
    ATET      & 0.009 & 0.025 & -0.063** & -0.117*** & -0.082*** \\
              & (0.010) & (0.018) & (0.026) & (0.024) & (0.026) \\
    Observations & 3,736,506 & 3,325,175 & 3,736,506 & 2,757,966 & 2,555,554 \\
    Controls  & YES & YES & YES & YES & YES \\
    \bottomrule
    \end{tabular}
    \begin{tablenotes}[para,flushleft]
\footnotesize{The table reports results following the difference-in-difference approach by \citet{Callaway_Santanna_2021}. ATET coefficients are obtained as a weighted average that considers the importance of each cohort of firms. The estimator is doubly robust, and the matching is obtained by controlling for firm size, age, capital intensity, TFP and 2-digit industry. There are 2,517 firms in the treated group for vertical integrations at 25th percentile and 956 treated firms for vertical integrations at the 75th percentile. The control group includes firms that have never been treated and those that have not been treated yet. Standard errors clustered at the firm level are in parentheses.  *** p$<$0.01, ** p$<$0.05, * p$<$0.1}.
\end{tablenotes}
\end{threeparttable}
  \end{subtable}

\end{table}

\subsection{Sample compositione effects}

A main concern that we have, relates to sample composition, as the takeovers have a diverse industrial and geographical coverage. In the first case, we want to check whether heterogeneity in our findings can emerge once we separate industries with an implicit similar level of technology intensity in the production process. In fact, we know from previous industrial organization literature \citep{berry_2019} that technology impacts market structures. Based on firms' industrial affiliations, we perform an exercise to classify subsidiaries in Low, Medium-Low, Medium-High, and High technological intensity. The classification is sourced from Eurostat and is based on the sector-level amount of research and development expenses and on the propensity to generate intellectual property rights. Appendix Table \ref{tab:tech_distribution} reports sample coverage along this dimension, showing that almost half of the target firms are active in low-tech industries, while High-Tech represents just 3\% of the sample. Eventually, we estimate the impact of the acquisition on each subsample using our baseline methodology explained in Section \ref{sec: strategy}. As shown in Table \ref{tab:tech_vertical_b}, the negative impact on markups after vertical strategies on supply chains is mainly explained by the Medium-High tech and Medium-Low tech industries. The other categories do not show any statistical significance on \textit{ex-post} markups.
We argue that the latter evidence is consistent with the intuition that there is more room to reduce profit margins in technological production processes whose markups are relatively higher than in low-tech production processes. The rationale behind this is that markups are higher in companies that tend to make substantial investments in innovation and R\&D, leading to higher prices. On the contrary, low-tech productions, in general, leave little room for markups and profit margins, while firms in High-tech industries may have oligopolistic or monopolist market structures that are less responsive to changes in ownership. Please note, however, once again that our estimates never point to positive impacts on markups after takeovers, and that is highly counterintuitive, especially in those technological sectors where one would expect an incentive to raise prices to recover the relevant R\&D costs.


\begin{table}[H]
  \centering
  \small
  \caption{Average treatment effect on the treated (ATET): classification by technology intensity after vertical integration}\label{tab:tech_vertical_a}
  \begin{subtable}[t]{\linewidth}
  \footnotesize
    \centering
    \vspace{0pt}
    \caption{Low Tech}
    \resizebox{\columnwidth}{!}{%
    \begin{tabular}{lcccccc} \hline
  	&	(1)	&	(2)	&	(3)	&	(4)	&	(5)	\\
VARIABLES 	&	Markup	&	Market share &	Sales &	Variable cost &	\makecell[c]{Variable cost \\ ratio} \\ \hline
ATET 	&	0.0004	&	0.035	&	-0.001	&	0.001	&	0.009	\\
&	(0.005)	&	(0.024)	&	(0.017)	&	(0.019)	&	(0.006)	\\
Observations &	1,623,719	&	1,623,742	&	1,623,742	&	1,623,742	&	1,614,590	\\ \hline
& (6) &	(7)	& (8) & (9) & (10) \\ 
VARIABLES &	TFP	&	ROI & \makecell[c]{Capital \\ Intensity} & \makecell[c]{Liquidity \\ Ratio} & \makecell[c]{Solvency \\ Ratio} \\ \hline
ATET &	0.015	&	0.015	&	-0.077**	&	-0.068**	&	-0.047	\\
&	(0.014)	&	(0.018)	&	(0.033)	&	(0.030)	&	(0.031)	\\
Observations &	1,623,742	&	1,437,506	&	1,623,742	&	1,209,443	&	1,094,396	\\
\hline \\
\end{tabular}
}
  \end{subtable}
  \begin{subtable}[t]{\linewidth}
    \centering
     \footnotesize
    \vspace{0pt}
     \resizebox{\columnwidth}{!}{%
    \begin{threeparttable}
    \caption{Medium-low Tech}
    \begin{tabular}{lccccccccc} \hline
  	&	(1)	&	(2)	&	(3)	&	(4)	&	(5)	\\
VARIABLES 	&	Markup	&	Market share	&	Sales	&	Variable cost	&	\makecell[c]{Variable cost \\ ratio} \\ \hline
ATET 	&	-0.007*	&	0.033	&	0.037**	&	0.043**	&	0.006	\\
&	(0.003)	&	(0.023)	&	(0.017)	&	(0.019)	&	(0.006)	\\
Observations &	1,321,184	&	1,321,188	&	1,321,188	&	1,321,188	&	1,314,033	\\ \hline 
& (6)	&	(7)	& (8) & (9) & (10) \\ 
VARIABLES &	TFP	&	ROI & \makecell[c]{Capital \\ Intensity} & \makecell[c]{Liquidity \\ Ratio} & \makecell[c]{Solvency \\ Ratio} \\ \hline
ATET &	-0.013	&	0.027	&	-0.047	&	-0.004	&	-0.063*	\\
&	(0.010)	&	(0.022)	&	(0.034)	&	(0.031)	&	(0.036)	\\
Observations &	1,321,188	&	1,178,573	&	1,321,188	&	972,005	&	909,863	\\
\hline
\end{tabular}
 \begin{tablenotes}[para,flushleft]
\footnotesize{The table shows results after the doubly robust estimator proposed by \citet{Callaway_Santanna_2021}, using never-treated and not-yet-treated units in the control group. Variables are in logs. There are 576 treated firms in the low-tech sample and 567 treated firms in the medium-low tech. Control variables are included but not reported. Standard errors clustered at the firm level are reported in parentheses, and significance levels are *** p$<$0.01 ** p$<$0.05 * p$<$0.1}
\end{tablenotes}
\end{threeparttable} 
  }  
  \end{subtable}
\end{table}


  \begin{table}[H]
  \centering
  \small
  \caption{Average treatment effect on the treated (ATET): classification by technology intensity after vertical integration}\label{tab:tech_vertical_b}
  \begin{subtable}[t]{\linewidth}
    \centering
     \footnotesize
    \vspace{0pt}
    \caption{Medium-high Tech}
    \resizebox{\columnwidth}{!}{%
    \begin{tabular}{lccccccccc} \hline
  	&	(1)	&	(2)	&	(3)	&	(4)	&	(5)	\\
VARIABLES 	&	Markup	&	Market share	&	Sales	&	Variable cost	&	\makecell[c]{Variable cost \\ ratio} \\ \hline
ATET 	&	-0.012**	&	0.033	&	0.044***	&	0.047***	&	0.005	\\
&	(0.005)	&	(0.024)	&	(0.017)	&	(0.017)	&	(0.008)	\\
Observations &	686,904	&	686,908	&	686,908	&	686,908	&	683,877	\\ \hline
& (6)	&	(7)	& (8) & (9) & (10) \\ 
VARIABLES &	TFP	& ROI & \makecell[c]{Capital \\ Intensity} & \makecell[c]{Liquidity \\ Ratio} & \makecell[c]{Solvency \\ Ratio} \\ \hline
ATET &	0.007	&	0.029	&	-0.083**	&	-0.198***	&	-0.239***	\\
&	(0.016)	&	(0.024)	&	(0.032)	&	(0.029)	&	(0.043)	\\
Observations &	686,908	&	615,833	&	686,908	&	497,556	&	475,951	\\
\hline \\
\end{tabular}
}
  \end{subtable}
  \begin{subtable}[t]{\linewidth}
    \centering
     \footnotesize
    \caption{High Tech}
    \resizebox{\columnwidth}{!}{%
    \begin{threeparttable}
    \footnotesize
    \begin{tabular}{lccccccccc} \hline
 	&	(1)	&	(2)	&	(3)	&	(4)	&	(5)	\\
VARIABLES 	&	Markup	&	Market share	&	Sales	&	Variable cost	&	\makecell[c]{Variable cost \\ ratio} \\ \hline
ATET 	&	-0.013	&	-0.027	&	0.017	&	0.011	&	-0.038	\\
&	(0.015)	&	(0.048)	&	(0.035)	&	(0.038)	&	(0.035)	\\
Obervations &	115,732	&	115,734	&	115,734	&	115,734	&	115,016	\\ \hline
& (6)	&	(7)	& (8) & (9) & (10) \\ 
VARIABLES &	TFP	&	ROI & \makecell[c]{Capital \\ Intensity} & \makecell[c]{Liquidity \\ Ratio} & \makecell[c]{Solvency \\ Ratio} \\ \hline
ATET &	0.008	&	-0.000	&	-0.029	&	-0.045	&	-0.078	\\
&	(0.019)	&	(0.053)	&	(0.071)	&	(0.057)	&	(0.056)	\\
&	115,734	&	103,608	&	115,734	&	85,983	&	82,262	\\
       \hline
    \end{tabular}
    \begin{tablenotes}[para,flushleft]
\footnotesize{The table shows results after the doubly robust estimator proposed by \citet{Callaway_Santanna_2021}, using never-treated and not-yet-treated units in the control group. Variables are in logs.  There are 669 in the medium-high tech sample and 143 treated firms in the high tech. Control variables are included but not reported. Standard errors clustered at the firm level are reported in parentheses and significance levels are *** p$<$0.01 ** p$<$0.05 * p$<$0.1}
\end{tablenotes}
\end{threeparttable}
    }
    
  \end{subtable}
\end{table}

\begin{figure}[H]
\caption{The case of EU-15 - two-period event studies on (log of) markups before and after takeovers}
\centerline{\includegraphics[scale=0.5]{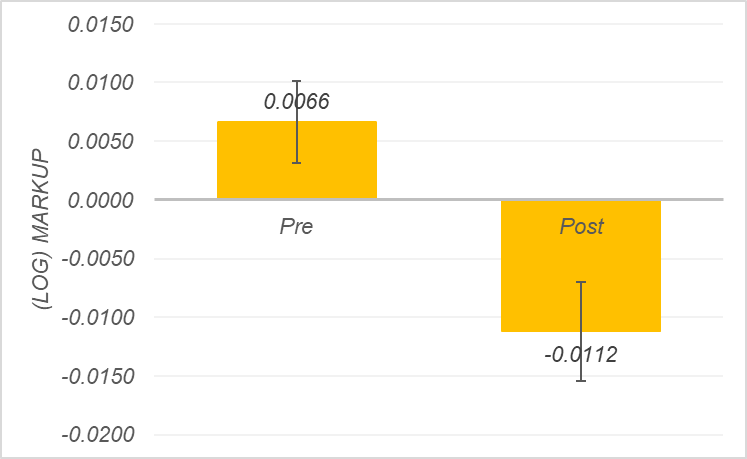}}
 \begin{tablenotes}
      \footnotesize
      \singlespacing
      \item Note: Pre- and post-treatment average markup changes estimated after a panel set up, following \citet{Callaway_Santanna_2021}. EU-15 include EU countries accessing before 2004. Bars indicate coefficients, while lines indicate 95\% confidence intervals.   
   \end{tablenotes}
\label{fig: eu15}
\end{figure}

\begin{figure}[H]
\caption{The case of BIG 5 EU members - two-period event studies on (log of) markups before and after takeovers}
\centerline{\includegraphics[scale=0.5]{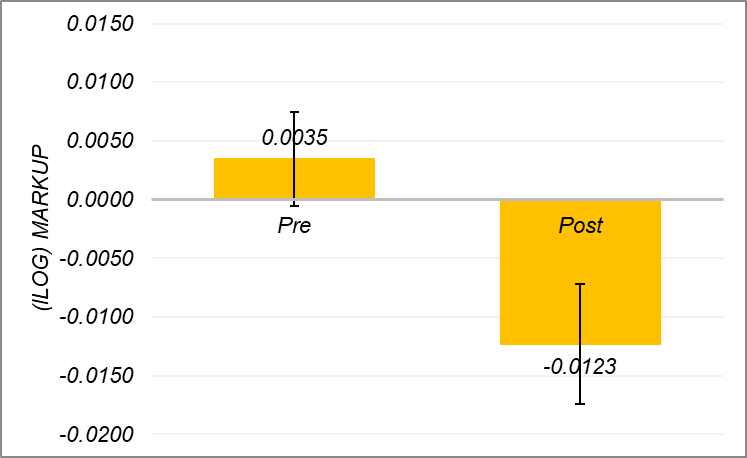}}
 \begin{tablenotes}
      \footnotesize
      \singlespacing
      \item Note: Pre- and post-treatment average markup changes estimated after a panel set up, following \citet{Callaway_Santanna_2021}. Big 5 EU members include Germany, France, Italy, Spain, and the United Kingdom. Bars indicate coefficients, while lines indicate 95\% confidence intervals. 
   \end{tablenotes}
\label{fig: big5}
\end{figure}

\begin{figure}[H]
\caption{The case foreign takeovers - two-period event studies on (log of) markups before and after takeovers}
\centerline{\includegraphics[scale=0.5]{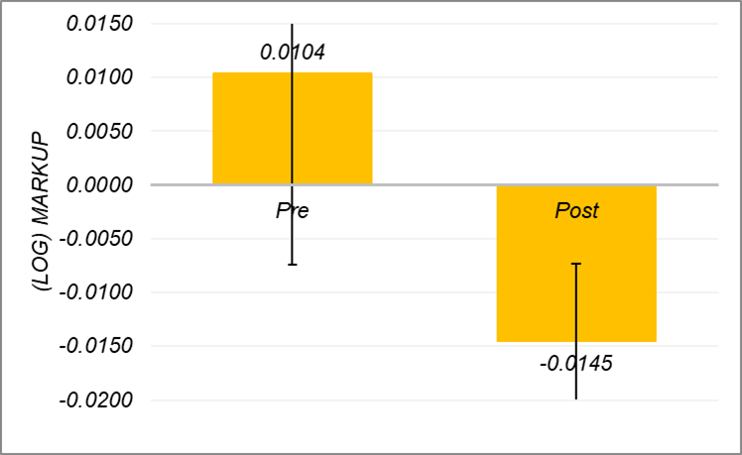}}
 \begin{tablenotes}
      \footnotesize
      \singlespacing
      \item Note: Pre- and post-treatment average markup changes estimated after a panel set up, following \citet{Callaway_Santanna_2021}. Foreign takeovers are defined as takeovers where the parent company is located in a country different from the target. Bars indicate coefficients, while lines indicate 95\% confidence intervals. 
   \end{tablenotes}
\label{fig: foreign}
\end{figure}

\newpage
In the next paragraphs, we will consider geographic heterogeneity. The first concern is that the European Union is not a fully integrated common market, and, for example, Member States that had recently accessed the Union could respond to different market incentives as they have a past as non-market economies. The intuition is that, in those Member States, markups can decrease because of a still ongoing transition to more competitive market structures, where prices go down with lower barriers to entry. If this is the case, our general results are biased by the presence of those firms that have been liberalized. Therefore, in Figure \ref{fig: eu15}, we visually reproduce the results of our baseline exercise with vertical integration strategies considering only the so-called EU-15, i.e. those countries that were Member States of the European Union before 2004. For the sake of simplicity, we report a two-period event study where bars represent the average changes in markups before and after takeovers\footnote{Please note that behind Figures \ref{fig: eu15}, \ref{fig: big5}, and \ref{fig: foreign}, we still perform a panel setup \'a la \citet{Callaway_Santanna_2021}, where we consider full cohorts of treated firms year by year, inverse probability weighting and a double robust estimator. The two-period event studies we report, thus, still consider different treatment timing and staggered effects.}. Clearly, we find that the impact on markups by takeovers is still negative, with an average magnitude (-1.1\%) higher than in the total sample.

In Figure \ref{fig: big5}, we further check whether results are robust when we limit our analyses only to the 5 biggest economies in the European Union (Germany, France, Italy, Spain, United Kingdom\footnote{As explained in the Data section, we keep the United Kingdom as a Member State of the European Union throughout our analyses because Brexit finally came to a withdrawal agreement only in 2020.}). In this case, the average changes in markups after vertical takeovers are about -1.2\%, therefore still higher than baseline results, and comparable to previous results on the sample of EU-15.

Finally, in Figure \ref{fig: foreign}, we separate the cases of foreign takeovers when the acquiring company is located in a country different from the target's. The concern is that multinational enterprises may have different pricing rules for international outsourcing when they can address a variety of markets. In our sample, foreign takeovers constitute about one-third of the total. Coefficients visualized in Figure \ref{fig: foreign} tell us that, on average, post-treatment markup changes are negative (-1.4\%) and significant. They are, on average, higher than previous baseline analyses.

\section{Conclusion}\label{sec: conclusions}
Rising market power has become a central concern for scholars and competition authorities, as it may indicate excessive market concentration and potential harm to consumer welfare. New tools to understand market concentration are especially useful after a revamped agenda for industrial policy by countries that aspire to create national champions. While most of the empirical studies of the literature are focused on what is happening in the United States, we still need more evidence about the European Union, where competition authorities have been historically more stringent on industrial aggregations and market abuses. 
After descriptive statistics on markups in the manufacturing industries of the European Union, we do observe an increasing trend, especially since 2012. Yet, when we investigate the causal impact of takeovers as one possible channel of increasing markups, we conclude that there is either no significant effect when firms pursue horizontal integration strategies, in the case of targets in the same industry, or there is a negative effect in the case of vertical integration strategies when the objective is to integrate buyers or suppliers on a supply chain. 
Therefore, we argue that our results point to the existence of  efficiency gains achieved through eliminating double profit margins. Eliminating double margins implies that a vertically integrated company can reduce the chain of successive markups along a supply chain, which would stand as a negative externality if independent companies exchange on the inputs markets. Thanks to lower markups, vertically integrated firms can increase their market shares and their sales, although we detect financial distress as liquidity and solvency ratios worsen after the acquisition.

An important result that goes beyond the textbook case of eliminating double margins is the negative causal relationship that we detect in our sample between the markups of the vertically integrated companies and the size of the corporate perimeters of the acquiring parents. Briefly, if the parent company has already integrated many subsidiaries, the negative impact on markups is relatively higher. The latter is a finding that points to bigger welfare efficiency gains in the presence of full segments of vertically integrated supply chain tasks.

Finally, our microeconometric results on firms' dynamics are apparently at odds with the evidence of an aggregate rising trend in market power detected at the continental level. Throughout our analyses, we never find that markups increase as a result of takeovers. Therefore, further investigations are needed to understand whether there are other channels different from takeovers that could explain the aggregate rising markups or perhaps general equilibrium effects are at work. A suggestion from policymakers in the most recent Vertical Merger Guidelines is that we should also consider possible foreclosure effects in inputs markets along supply chains. Unfortunately, our data and our empirical tools do not allow us to identify indirect effects on the market structure. Yet, we argue that the latter concerns do not undermine the importance of considering direct efficiency gains brought about by vertically integrated supply chains, which deserve more attention by policymakers and scholars who want to understand the whys and wherefores of global market power.

\bibliographystyle{chicago}
\bibliography{bib_paper}

\begin{thebibliography}{}

\bibitem[\protect\citeauthoryear{Acemoglu, Johnson, and Mitton}{Acemoglu et~al.}{2009}]{acemoglu_2009}
Acemoglu, D., S.~Johnson, and T.~Mitton (2009).
\newblock Determinants of vertical integration: financial development and contracting costs.
\newblock {\em The journal of finance\/}~{\em 64\/}(3), 1251--1290.

\bibitem[\protect\citeauthoryear{Ackerberg, Caves, and Frazer}{Ackerberg et~al.}{2015}]{Ackerberg_Caves_Frazer_2015}
Ackerberg, D.~A., K.~Caves, and G.~Frazer (2015).
\newblock Identification properties of recent production function estimators.
\newblock {\em Econometrica\/}~{\em 83\/}(6), 2411--2451.

\bibitem[\protect\citeauthoryear{Alfaro, Conconi, Fadinger, and Newman}{Alfaro et~al.}{2016}]{alfaro_2016}
Alfaro, L., P.~Conconi, H.~Fadinger, and A.~F. Newman (2016).
\newblock Do prices determine vertical integration?
\newblock {\em The Review of Economic Studies\/}~{\em 83\/}(3), 855--888.

\bibitem[\protect\citeauthoryear{Alviarez, Head, and Mayer}{Alviarez et~al.}{2025}]{alviarez_2020}
Alviarez, V., K.~Head, and T.~Mayer (2025).
\newblock Global giants and local stars: How changes in brand ownership affect competition.
\newblock {\em American Economic Journal: Microeconomics\/}~{\em 17\/}(1), 389--432.

\bibitem[\protect\citeauthoryear{Antr{\`a}s}{Antr{\`a}s}{2020}]{antras_2020}
Antr{\`a}s, P. (2020).
\newblock Conceptual aspects of global value chains.
\newblock {\em The World Bank Economic Review\/}~{\em 34\/}(3), 551--574.

\bibitem[\protect\citeauthoryear{Atalay, Horta{\c{c}}su, and Syverson}{Atalay et~al.}{2014}]{atalay_2014}
Atalay, E., A.~Horta{\c{c}}su, and C.~Syverson (2014).
\newblock Vertical integration and input flows.
\newblock {\em American Economic Review\/}~{\em 104\/}(4), 1120--48.

\bibitem[\protect\citeauthoryear{Autor, Dorn, Katz, Patterson, and Van~Reenen}{Autor et~al.}{2020}]{Autor_et_al_2020}
Autor, D., D.~Dorn, L.~F. Katz, C.~Patterson, and J.~Van~Reenen (2020).
\newblock {The Fall of the Labor Share and the Rise of Superstar Firms*}.
\newblock {\em The Quarterly Journal of Economics\/}~{\em 135\/}(2), 645--709.

\bibitem[\protect\citeauthoryear{Baker, Rose, Salop, and Morton}{Baker et~al.}{2018}]{baker_2018}
Baker, J.~B., N.~L. Rose, S.~C. Salop, and F.~S. Morton (2018).
\newblock Five principals for vertical merger enforcement policy.
\newblock {\em Antitrust\/}~{\em 33}, 12.

\bibitem[\protect\citeauthoryear{Baqaee and Farhi}{Baqaee and Farhi}{2020}]{baqaee_2020}
Baqaee, D.~R. and E.~Farhi (2020).
\newblock Productivity and misallocation in general equilibrium.
\newblock {\em The Quarterly Journal of Economics\/}~{\em 135\/}(1), 105--163.

\bibitem[\protect\citeauthoryear{Basu}{Basu}{2019}]{Basu_2019}
Basu, S. (2019).
\newblock Are price-cost markups rising in the united states? a discussion of the evidence.
\newblock {\em The Journal of Economic Perspectives\/}~{\em 33\/}(3), 3--22.

\bibitem[\protect\citeauthoryear{Berry, Gaynor, and Scott~Morton}{Berry et~al.}{2019}]{berry_2019}
Berry, S., M.~Gaynor, and F.~Scott~Morton (2019).
\newblock Do increasing markups matter? lessons from empirical industrial organization.
\newblock {\em Journal of Economic Perspectives\/}~{\em 33\/}(3), 44--68.

\bibitem[\protect\citeauthoryear{Berto Villas-Boas}{Berto Villas-Boas}{2007}]{Berto_2007}
Berto Villas-Boas, S. (2007).
\newblock Vertical relationships between manufacturers and retailers: Inference with limited data.
\newblock {\em The Review of Economic Studies\/}~{\em 74\/}(2), 625--652.

\bibitem[\protect\citeauthoryear{Bertrand and Zitouna}{Bertrand and Zitouna}{2008}]{Bertrand_Zitouna_2008}
Bertrand, O. and H.~Zitouna (2008).
\newblock Domestic versus cross-border acquisitions: which impact on the target firms’ performance?
\newblock {\em Applied Economics\/}~{\em 40\/}(17), 2221--2238.

\bibitem[\protect\citeauthoryear{Bighelli, Di~Mauro, Melitz, and Mertens}{Bighelli et~al.}{2023}]{Bighelli_2023}
Bighelli, T., F.~Di~Mauro, M.~J. Melitz, and M.~Mertens (2023).
\newblock European firm concentration and aggregate productivity.
\newblock {\em Journal of the European Economic Association\/}~{\em 21\/}(2), 455--483.

\bibitem[\protect\citeauthoryear{Blonigen and Pierce}{Blonigen and Pierce}{2016}]{Blonigen_2016}
Blonigen, B.~A. and J.~R. Pierce (2016).
\newblock Evidence for the effects of mergers on market power and efficiency.
\newblock Technical report, National Bureau of Economic Research.

\bibitem[\protect\citeauthoryear{Bond, Hashemi, Kaplan, and Zoch}{Bond et~al.}{2021}]{bond_2021}
Bond, S., A.~Hashemi, G.~Kaplan, and P.~Zoch (2021).
\newblock Some unpleasant markup arithmetic: Production function elasticities and their estimation from production data.
\newblock {\em Journal of Monetary Economics\/}~{\em 121}, 1--14.

\bibitem[\protect\citeauthoryear{Callaway and Sant’Anna}{Callaway and Sant’Anna}{2021}]{Callaway_Santanna_2021}
Callaway, B. and P.~H. Sant’Anna (2021).
\newblock Difference-in-differences with multiple time periods.
\newblock {\em Journal of econometrics\/}~{\em 225\/}(2), 200--230.

\bibitem[\protect\citeauthoryear{Chipty}{Chipty}{2001}]{chipty_2001}
Chipty, T. (2001).
\newblock Vertical integration, market foreclosure, and consumer welfare in the cable television industry.
\newblock {\em American Economic Review\/}~{\em 91\/}(3), 428--453.

\bibitem[\protect\citeauthoryear{Choné, Linnemer, and Vergé}{Choné et~al.}{2023}]{chone_2023}
Choné, P., L.~Linnemer, and T.~Vergé (2023).
\newblock Double marginalization, market foreclosure, and vertical integration.
\newblock {\em Journal of the European Economic Association\/}~{\em 22\/}(4), 1884--1935.

\bibitem[\protect\citeauthoryear{Ciliberto}{Ciliberto}{2006}]{ciliberto_2006}
Ciliberto, F. (2006).
\newblock Does organizational form affect investment decisions?
\newblock {\em The journal of industrial economics\/}~{\em 54\/}(1), 63--93.

\bibitem[\protect\citeauthoryear{Comanor}{Comanor}{1967}]{comanor_1967}
Comanor, W.~S. (1967).
\newblock Vertical mergers, market powers, and the antitrust laws.
\newblock {\em The American economic review\/}~{\em 57\/}(2), 254--265.

\bibitem[\protect\citeauthoryear{{Competition and Markets Authority}}{{Competition and Markets Authority}}{2021}]{UK2021}
{Competition and Markets Authority} (2021).
\newblock Merger assessment guidelines, cma129.

\bibitem[\protect\citeauthoryear{Cravino and Levchenko}{Cravino and Levchenko}{2016}]{Cravino_Levchenko_2016}
Cravino, J. and A.~A. Levchenko (2016).
\newblock {Multinational Firms and International Business Cycle Transmission*}.
\newblock {\em The Quarterly Journal of Economics\/}~{\em 132\/}(2), 921--962.

\bibitem[\protect\citeauthoryear{Crawford, Lee, Whinston, and Yurukoglu}{Crawford et~al.}{2018}]{crawford_2018}
Crawford, G.~S., R.~S. Lee, M.~D. Whinston, and A.~Yurukoglu (2018).
\newblock The welfare effects of vertical integration in multichannel television markets.
\newblock {\em Econometrica\/}~{\em 86\/}(3), 891--954.

\bibitem[\protect\citeauthoryear{De~Loecker et~al.}{De~Loecker et~al.}{2021}]{DeLoecker_2021}
De~Loecker, J. et~al. (2021).
\newblock Comment on (un) pleasant... by bond et al (2020).
\newblock {\em Journal of Monetary Economics\/}~{\em 121\/}(C), 15--18.

\bibitem[\protect\citeauthoryear{De~Loecker and Eeckhout}{De~Loecker and Eeckhout}{2018}]{DeLoecker_Eeckhout_2018}
De~Loecker, J. and J.~Eeckhout (2018).
\newblock Global market power.
\newblock Working Paper 24768, National Bureau of Economic Research.

\bibitem[\protect\citeauthoryear{De~Loecker, Eeckhout, and Unger}{De~Loecker et~al.}{2020}]{DeLoecker_Eeckhout_2020}
De~Loecker, J., J.~Eeckhout, and G.~Unger (2020).
\newblock {The Rise of Market Power and the Macroeconomic Implications*}.
\newblock {\em The Quarterly Journal of Economics\/}~{\em 135\/}(2), 561--644.

\bibitem[\protect\citeauthoryear{De~Loecker and Warzynski}{De~Loecker and Warzynski}{2012}]{DeLoecker_Warzynski_2012}
De~Loecker, J. and F.~Warzynski (2012).
\newblock Markups and firm-level export status.
\newblock {\em The American Economic Review\/}~{\em 102\/}(6), 2437--2471.

\bibitem[\protect\citeauthoryear{Deb, Eeckhout, Patel, and Warren}{Deb et~al.}{2022}]{deb_2022}
Deb, S., J.~Eeckhout, A.~Patel, and L.~Warren (2022).
\newblock What drives wage stagnation: Monopsony or monopoly?
\newblock {\em Journal of the European Economic Association\/}~{\em 20\/}(6), 2181--2225.

\bibitem[\protect\citeauthoryear{{Del Prete} and Rungi}{{Del Prete} and Rungi}{2017}]{DelPrete_Rungi_2017}
{Del Prete}, D. and A.~Rungi (2017).
\newblock Organizing the global value chain: A firm-level test.
\newblock {\em Journal of International Economics\/}~{\em 109}, 16--30.

\bibitem[\protect\citeauthoryear{{Del Prete} and Rungi}{{Del Prete} and Rungi}{2020}]{DelPrete_Rungi_2020}
{Del Prete}, D. and A.~Rungi (2020).
\newblock Backward and forward integration along global value chains.
\newblock {\em Review of Industrial Organization\/}~{\em 57}, 263--283.

\bibitem[\protect\citeauthoryear{Dewulf, Klein, Mell, and Shchepetova}{Dewulf et~al.}{2022}]{Dewulf_et_al_2022}
Dewulf, S., T.~Klein, A.~Mell, and A.~Shchepetova (2022).
\newblock {EU and UK Vertical Merger Control: What’s the State of Play?}
\newblock {\em Journal of European Competition Law and Practice\/}~{\em 14\/}(2), 113--120.

\bibitem[\protect\citeauthoryear{Duran-Micco and Perloff}{Duran-Micco and Perloff}{2022}]{duran_2022}
Duran-Micco, E. and J.~M. Perloff (2022).
\newblock How large are double markups?
\newblock {\em International Journal of Industrial Organization\/}~{\em 85}, 102885.

\bibitem[\protect\citeauthoryear{Díez, Fan, and Villegas-Sánchez}{Díez et~al.}{2021}]{Diez_at_al2021}
Díez, F.~J., J.~Fan, and C.~Villegas-Sánchez (2021).
\newblock Global declining competition?
\newblock {\em Journal of International Economics\/}~{\em 132}, 103492.

\bibitem[\protect\citeauthoryear{{European Commission}}{{European Commission}}{2008}]{EU2008}
{European Commission} (2008).
\newblock Guidelines on the assessment of non-horizontal mergers under the council regulation on the control of concentrations between undertakings’, 2008/c 265/07.
\newblock Technical report, European Commission.

\bibitem[\protect\citeauthoryear{Eurostat}{Eurostat}{2007}]{eurostat2007for}
Eurostat (2007).
\newblock Reccomendations {Manual} on the {Production} of {Foreign Affiliates Statistics}.
\newblock {\em Eurostat Methodologies and working papers\/}.

\bibitem[\protect\citeauthoryear{Fan and Lang}{Fan and Lang}{2000}]{fan_2000}
Fan, J.~P. and L.~H. Lang (2000).
\newblock The measurement of relatedness: An application to corporate diversification.
\newblock {\em The Journal of Business\/}~{\em 73\/}(4), 629--660.

\bibitem[\protect\citeauthoryear{{Federal Trade Commission}}{{Federal Trade Commission}}{2020}]{USguidelines2020}
{Federal Trade Commission} (2020).
\newblock {US} vertical merger guidelines.
\newblock Technical report, Federal Trade Commission.

\bibitem[\protect\citeauthoryear{{Federal Trade Commission}}{{Federal Trade Commission}}{2023}]{USguidelines2023}
{Federal Trade Commission} (2023).
\newblock {US} vertical merger guidelines.
\newblock Technical report, Federal Trade Commission.

\bibitem[\protect\citeauthoryear{Gil}{Gil}{2015}]{Gil_2015}
Gil, R. (2015).
\newblock Does vertical integration decrease prices? evidence from the paramount antitrust case of 1948.
\newblock {\em American Economic Journal: Economic Policy\/}~{\em 7\/}(2), 162--91.

\bibitem[\protect\citeauthoryear{Gil and Warzynski}{Gil and Warzynski}{2015}]{gil_warz_2015}
Gil, R. and F.~Warzynski (2015).
\newblock Vertical integration, exclusivity, and game sales performance in the us video game industry.
\newblock {\em The journal of law, economics, and organization\/}~{\em 31\/}(suppl\_1), i143--i168.

\bibitem[\protect\citeauthoryear{Grassi, De~Ridder, and Morzenti}{Grassi et~al.}{2022}]{Basile_et_al_2024}
Grassi, B., M.~De~Ridder, and G.~Morzenti (2022).
\newblock The hitchhiker's guide to markup estimation: Assessing estimates from financial data.
\newblock CEPR Discussion Paper No. 17532.
\newblock unpublished new version 2024.

\bibitem[\protect\citeauthoryear{Grieco, Pinkse, and Slade}{Grieco et~al.}{2018}]{grieco_2018}
Grieco, P., J.~Pinkse, and M.~Slade (2018).
\newblock Brewed in north america: Mergers, marginal costs, and efficiency.
\newblock {\em International Journal of Industrial Organization\/}~{\em 59}, 24--65.

\bibitem[\protect\citeauthoryear{Grullon, Larkin, and Michaely}{Grullon et~al.}{2019}]{Grullon_et_al_2019}
Grullon, G., Y.~Larkin, and R.~Michaely (2019).
\newblock {Are US Industries Becoming More Concentrated?}
\newblock {\em Review of Finance\/}~{\em 23\/}(4), 697--743.

\bibitem[\protect\citeauthoryear{Gugler, Mueller, Yurtoglu, and Zulehner}{Gugler et~al.}{2003}]{Gugler_et_al_2003}
Gugler, K., D.~C. Mueller, B.~Yurtoglu, and C.~Zulehner (2003).
\newblock The effects of mergers: an international comparison.
\newblock {\em International Journal of Industrial Organization\/}~{\em 21\/}(5), 625--653.

\bibitem[\protect\citeauthoryear{Gutierrez and Philippon}{Gutierrez and Philippon}{2023}]{Gutierrez_Philippon_2023}
Gutierrez, G. and T.~Philippon (2023).
\newblock How european markets became free: A study of institutional drift.
\newblock {\em Journal of the European Economic Association\/}~{\em 21\/}(1), 251--292.

\bibitem[\protect\citeauthoryear{Hall}{Hall}{1988}]{Hall_1988}
Hall, R.~E. (1988).
\newblock The relation between price and marginal cost in u.s. industry.
\newblock {\em Journal of Political Economy\/}~{\em 96\/}(5), 921--947.

\bibitem[\protect\citeauthoryear{Hall}{Hall}{2018}]{Hall_2018}
Hall, R.~E. (2018).
\newblock New evidence on the markup of prices over marginal costs and the role of mega-firms in the us economy.
\newblock Technical report, National Bureau of Economic Research.

\bibitem[\protect\citeauthoryear{Hastings and Gilbert}{Hastings and Gilbert}{2005}]{hastings_2005}
Hastings, J.~S. and R.~J. Gilbert (2005).
\newblock Market power, vertical integration and the wholesale price of gasoline.
\newblock {\em The Journal of Industrial Economics\/}~{\em 53\/}(4), 469--492.

\bibitem[\protect\citeauthoryear{Horta{\c{c}}su and Syverson}{Horta{\c{c}}su and Syverson}{2007}]{Hortaccsu_Syverson_2007}
Horta{\c{c}}su, A. and C.~Syverson (2007).
\newblock Cementing relationships: Vertical integration, foreclosure, productivity, and prices.
\newblock {\em Journal of political economy\/}~{\em 115\/}(2), 250--301.

\bibitem[\protect\citeauthoryear{Karlinger, Magos, and Zenger}{Karlinger et~al.}{2020}]{Karlinger_et_al_2020}
Karlinger, L., D.~Magos, and H.~Zenger (2020).
\newblock Recent developments at dg competition: 2019/2020.
\newblock {\em eview of Industrial Organization\/}~{\em 57\/}(4), 783--814.

\bibitem[\protect\citeauthoryear{Kwoka and Slade}{Kwoka and Slade}{2019}]{kwoka_2019}
Kwoka, J. and M.~Slade (2019).
\newblock Second thoughts on double marginalization.
\newblock {\em Antitrust\/}~{\em 34}, 51.

\bibitem[\protect\citeauthoryear{Lafontaine and Slade}{Lafontaine and Slade}{2007}]{lafontaine_2007}
Lafontaine, F. and M.~Slade (2007).
\newblock Vertical integration and firm boundaries: The evidence.
\newblock {\em Journal of Economic literature\/}~{\em 45\/}(3), 629--685.

\bibitem[\protect\citeauthoryear{Luco and Marshall}{Luco and Marshall}{2020}]{luco_2020}
Luco, F. and G.~Marshall (2020).
\newblock The competitive impact of vertical integration by multiproduct firms.
\newblock {\em American Economic Review\/}~{\em 110\/}(7), 2041--64.

\bibitem[\protect\citeauthoryear{Maksimovic, Phillips, and Prabhala}{Maksimovic et~al.}{2011}]{Maksimovic_et_al_2011}
Maksimovic, V., G.~Phillips, and N.~Prabhala (2011).
\newblock Post-merger restructuring and the boundaries of the firm.
\newblock {\em Journal of Financial Economics\/}~{\em 102\/}(2), 317--343.

\bibitem[\protect\citeauthoryear{McAdam, Petroulakis, Vansteenkiste, Cavalleri, Eliet, and Soares}{McAdam et~al.}{2019}]{McAdam_et_al_2019}
McAdam, P., F.~Petroulakis, I.~Vansteenkiste, M.~C. Cavalleri, A.~Eliet, and A.~Soares (2019).
\newblock {Concentration, market power and dynamism in the euro area}.
\newblock Working Paper Series 2253, European Central Bank.

\bibitem[\protect\citeauthoryear{McGuckin and Nguyen}{McGuckin and Nguyen}{1995}]{McGukin_Nguyen_1995}
McGuckin, R.~H. and S.~V. Nguyen (1995).
\newblock On productivity and plant ownership change: New evidence from the longitudinal research database.
\newblock {\em The RAND Journal of Economics\/}~{\em 26\/}(2), 257--276.

\bibitem[\protect\citeauthoryear{Miller}{Miller}{2025}]{Miller_2024}
Miller, N.~H. (2025).
\newblock Industrial organization and the rise of market power.
\newblock {\em International Journal of Industrial Organization\/}~{\em 98}, 103131.

\bibitem[\protect\citeauthoryear{Morlacco}{Morlacco}{2019}]{morlacco_2019}
Morlacco, M. (2019).
\newblock Market power in input markets: Theory and evidence from french manufacturing.
\newblock {\em Unpublished, March\/}~{\em 20}, 2019.

\bibitem[\protect\citeauthoryear{OECD}{OECD}{2005}]{oecd2005mne}
OECD (2005).
\newblock Guidelines for {M}ultinational {E}nterprises.
\newblock {\em OECD Publishing\/}.

\bibitem[\protect\citeauthoryear{Raval}{Raval}{2023}]{raval_2023}
Raval, D. (2023).
\newblock Testing the production approach to markup estimation.
\newblock {\em Review of Economic Studies\/}~{\em 90\/}(5), 2592--2611.

\bibitem[\protect\citeauthoryear{Rosenbaum and Rubin}{Rosenbaum and Rubin}{1983}]{RosenbaumRubin1}
Rosenbaum, P.~R. and D.~B. Rubin (1983).
\newblock {The central role of the propensity score in observational studies for causal effects}.
\newblock {\em Biometrika\/}~{\em 70\/}(1), 41--55.

\bibitem[\protect\citeauthoryear{Rosenbaum and Rubin}{Rosenbaum and Rubin}{1985}]{RosenbaumRubin2}
Rosenbaum, P.~R. and D.~B. Rubin (1985).
\newblock Constructing a control group using multivariate matched sampling methods that incorporate the propensity score.
\newblock {\em The American Statistician\/}~{\em 39\/}(1), 33--38.

\bibitem[\protect\citeauthoryear{Rubens}{Rubens}{2023}]{rubens_2023}
Rubens, M. (2023).
\newblock Market structure, oligopsony power, and productivity.
\newblock {\em American Economic Review\/}~{\em 113\/}(9), 2382--2410.

\bibitem[\protect\citeauthoryear{Rubin}{Rubin}{2001}]{Rubin}
Rubin, D.~B. (2001).
\newblock Using propensity scores to help design observational studies: Application to the tobacco litigation.
\newblock {\em Health Services and Outcomes Research Methodology\/}~{\em 2\/}(3), 169--188.

\bibitem[\protect\citeauthoryear{Salinger}{Salinger}{1991}]{salinger_1991}
Salinger, M.~A. (1991).
\newblock Vertical mergers in multi-product industries and edgeworth's paradox of taxation.
\newblock {\em The journal of industrial economics\/}, 545--556.

\bibitem[\protect\citeauthoryear{Salop}{Salop}{2018}]{salop_2018}
Salop, S.~C. (2018).
\newblock Invigorating vertical merger enforcement.
\newblock {\em The Yale Law Journal\/}, 1962--1994.

\bibitem[\protect\citeauthoryear{Sant’Anna and Zhao}{Sant’Anna and Zhao}{2020}]{Sant'Anna_Zhao_2020}
Sant’Anna, P.~H. and J.~Zhao (2020).
\newblock Doubly robust difference-in-differences estimators.
\newblock {\em Journal of Econometrics\/}~{\em 219\/}(1), 101--122.

\bibitem[\protect\citeauthoryear{Slade}{Slade}{2021}]{slade_2021}
Slade, M.~E. (2021).
\newblock Vertical mergers: A survey of ex post evidence and ex ante evaluation methods.
\newblock {\em Review of Industrial Organization\/}~{\em 58\/}(4), 493--511.

\bibitem[\protect\citeauthoryear{Spengler}{Spengler}{1950}]{Spengler_1950}
Spengler, J.~J. (1950).
\newblock Vertical integration and antitrust policy.
\newblock {\em Journal of political economy\/}~{\em 58\/}(4), 347--352.

\bibitem[\protect\citeauthoryear{Stiebale and Vencappa}{Stiebale and Vencappa}{2018}]{Stiebale_Vencappa_2018}
Stiebale, J. and D.~Vencappa (2018).
\newblock Acquisitions, markups, efficiency, and product quality: Evidence from india.
\newblock {\em Journal of International Economics\/}~{\em 112}, 70--87.

\bibitem[\protect\citeauthoryear{Syverson}{Syverson}{2019}]{syverson_2019}
Syverson, C. (2019).
\newblock Macroeconomics and market power: Context, implications, and open questions.
\newblock {\em Journal of Economic Perspectives\/}~{\em 33\/}(3), 23--43.

\bibitem[\protect\citeauthoryear{Traina}{Traina}{2018}]{traina_2018}
Traina, J. (2018).
\newblock {\em Is Aggregate Market Power Increasing?: Production Trends Using Financial Statements}.
\newblock New working paper series. George J. Stigler Center for the Study of the Economy and the State, University of Chicago Booth School of Business.

\bibitem[\protect\citeauthoryear{UNCTAD}{UNCTAD}{2009}]{unctad2009fdi}
UNCTAD (2009).
\newblock Training manual on statistics for {FDI} and the operations of {TNC}s.
\newblock {\em United Nations\/}.

\bibitem[\protect\citeauthoryear{UNCTAD}{UNCTAD}{2016}]{unctad2016inv}
UNCTAD (2016).
\newblock World investment report 2016. {I}nvestor nationality: Policy challenges.
\newblock {\em United Nations\/}.

\bibitem[\protect\citeauthoryear{Van~Beveren}{Van~Beveren}{2012}]{van_2012}
Van~Beveren, I. (2012).
\newblock Total factor productivity estimation: A practical review.
\newblock {\em Journal of economic surveys\/}~{\em 26\/}(1), 98--128.

\bibitem[\protect\citeauthoryear{Van~Reenen}{Van~Reenen}{2018}]{VanReenen_2018}
Van~Reenen, J. (2018).
\newblock {Increasing differences between firms: market power and the macro-economy}.
\newblock Cep discussion papers, Centre for Economic Performance, LSE.

\end{thebibliography}

\clearpage
\appendix
\onehalfspacing

\section*{Appendix A: Markup estimation}\label{sec:appendix_markup}

\setcounter{table}{0}
\renewcommand{\thetable}{A\arabic{table}}
\setcounter{figure}{0}
\renewcommand{\thefigure}{A\arabic{figure}}

For firm-level markup estimation, we rely on the production function approach method proposed by \citet{DeLoecker_Warzynski_2012} and recently tested by \citet{raval_2023}, which assumes that firms are price-takers in inputs markets and they have continuous and twice differentiable production functions. Crucially, any firm $i$ uses at least one variable factor that can be freely adjusted each period $t$ after observing productivity shocks. In this context, the firm-level markup is the ratio of output price, $P_{it}$ over marginal cost, $MC_{it}$, because the output elasticity of the variable factor only equals its expenditure share in total revenue when prices equal the marginal costs of production. Hence, the presence of markups drives a wedge between the input's revenue share and its output elasticity. The empirical strategy relies on standard cost minimization conditions for variable inputs free of adjustment costs and on the estimation of output elasticity. 

In particular, given a production technology $Q_{it}=Q_{it}(X_{it}^1,...,X_{it}^V,K_{it},\omega_{it})$ with V variable inputs such as labor or intermediate inputs and assuming that producers are cost minimizers, the FOCs for any variable inputs associated with the Lagrangian function are such that:
\begin{equation*}
\frac{\partial L_{it}}{\partial X_{it}^{v}}= P_{it}^{X^{v}}-\lambda_{it}\frac{\partial Q_{it}(\cdot)}{\partial X_{it}^{v}}=0
\end{equation*}
where $\lambda_{it}$ is the marginal cost of production at a given level of output. Rearranging terms, multiplying both sides by $\frac{X_{it}}{Q_{it}}$ and defining $\mu_{it}\equiv \frac{P_{it}}{\lambda_{it}}$ the following expression for markups can be derived:
\begin{equation*}
\mu_{it}=\theta_{it}^{X} (\alpha_{it}^{X})^{-1}
\end{equation*}
where $\theta_{it}^{X}$ is the output elasticity on an in input X and $\alpha_{it}^{X}$ is the share of expenditures on input $X_{it}$ in total sales $(P_{it}Q_{it})$.
We estimate the output elasticity associated with a translog production function in which we use materials as a proxy for variable costs. To get estimates of the output elasticity, consider a production function with Hicks-neutral productivity term and common technology parameters across the set of producers:
\begin{equation*}
Q_{it}= F(X_{it}^{1},...,X_{it}^{V},K_{it};\beta)exp(\omega_{it})
\end{equation*}
This form allows us to rely on the proxy method suggested by \citet{Ackerberg_Caves_Frazer_2015} to obtain consistent estimates of the technology parameters $\beta$. The estimation procedure relies on using materials to proxy for productivity to solve the simultaneity problem deriving from unobserved productivity shocks potentially correlated with input choices. In particular, in the first stage, we run the following regression to obtain estimates of expected output $(\hat{\phi}_{it})$ and an estimate for $\epsilon_{it}$: $y_{it}=\phi_t(\ell_{it},k_{it},m_{it},\textbf{z}_{it})+\epsilon_{it}$, while in the second stage, we rely on the law of motion of productivity $\omega_{it}=g_t(\omega_{it-1})+\xi_{it}$ to get estimates for all production function coefficients. Table \ref{tab:profuction_function} shows the estimates of our revenue function coefficients disaggregated by industry. Note that the coefficient estimates we obtain are largely consistent with those reported in the literature (see, for example, \citet{van_2012}).


Finally, the advantage of the method employed compared with more standard methods in applied industrial organization \citep{Miller_2024} is that it requires significantly less data and assumptions and, therefore, can be applied to wider datasets of firm-level financial accounts. On the other hand, important pitfalls have been discussed by \citet{Basu_2019} and \citet{syverson_2019}. The recent work by \citet{bond_2021} highlights that identification issues arise when firm-level output prices are not observed. From this perspective, more traditional industry case studies are better at catching price effects, possibly with longer time series and detailed information on prices \citep{Miller_2024}. Yet, \citet{Basile_et_al_2024} demonstrates that revenue-based markups (like ours), calculated without detailed information on prices and changing demand, differ from true markups by a constant. Therefore, revenue-based markups from financial data and true (unobserved) markups have a correlation equal to one. Based on this finding, we can still investigate revenue-based markups in big financial datasets (like ours), discussing dispersions, growth rates, and the like, with however a \textit{caveat} in mind about their true magnitudes. 

\begin{table}[htbp]
\footnotesize
\caption {Production function coefficients by sectors}\label{tab:profuction_function}
\centering
\begin{threeparttable}
\begin{tabular}{cccc}
\small
Industry & $\beta^L$ & $\beta^M$ & $\beta^K$ \\
\hline
10 & 0.2034 & 0.6542 & 0.1129 \\
   & (0.1341) & (0.1673) & (0.0467) \\
11 & 0.2305 & 0.6904 & 0.0901 \\
   & (0.1029) & (0.1586) & (0.0491) \\
12 & 0.3546 & 0.6144 & 0.1260 \\
   & (0.1897) & (0.1884) & (0.0767) \\
13 & 0.3020 & 0.5576 & 0.1090 \\
   & (0.1489) & (0.2129) & (0.0515) \\
14 & 0.3279 & 0.4731 & 0.1103 \\
   & (0.1726) & (0.2737) & (0.0350) \\
15 & 0.3020 & 0.5138 & 0.1254 \\
   & (0.1628) & (0.2482) & (0.0472) \\
16 & 0.2084 & 0.6826 & 0.0812 \\
   & (0.1167) & (0.1678) & (0.0346) \\
17 & 0.2258 & 0.6726 & 0.0908 \\
   & (0.1181) & (0.1632) & (0.0342) \\
18 & 0.3774 & 0.4934 & 0.0858 \\
   & (0.1271) & (0.1791) & (0.0315) \\
19 & 0.2497 & 0.6302 & 0.1376 \\
   & (0.1582) & (0.2021) & (0.0872) \\
20 & 0.2733 & 0.6481 & 0.1006 \\
   & (0.1555) & (0.1713) & (0.0574) \\
21 & 0.3180 & 0.6068 & 0.1155 \\
   & (0.1697) & (0.1807) & (0.0720) \\
22 & 0.2227 & 0.6557 & 0.1016 \\
   & (0.1242) & (0.1642) & (0.0340) \\
23 & 0.2579 & 0.6421 & 0.0954 \\
   & (0.1129) & (0.1579) & (0.0432) \\
24 & 0.2862 & 0.6442 & 0.0695 \\
   & (0.1152) & (0.1566) & (0.0345) \\
25 & 0.3699 & 0.5133 & 0.1037 \\
   & (0.1294) & (0.1566) & (0.0214) \\
26 & 0.3384 & 0.5476 & 0.0917 \\
   & (0.1460) & (0.1957) & (0.0411) \\
27 & 0.2647 & 0.6220 & 0.0774 \\
   & (0.1384) & (0.1915) & (0.0381) \\
28 & 0.2825 & 0.6326 & 0.0663 \\
   & (0.1249) & (0.1731) & (0.0289) \\
29 & 0.2683 & 0.6371 & 0.0740 \\
   & (0.1519) & (0.1850) & (0.0401) \\
30 & 0.4275 & 0.4942 & 0.0811 \\
   & (0.1854) & (0.2112) & (0.0399) \\
31 & 0.1920 & 0.7135 & 0.0751 \\
   & (0.1242) & (0.1745) & (0.0326) \\
32 & 0.3069 & 0.5255 & 0.0964 \\
   & (0.0914) & (0.1552) & (0.0337) \\
\hline
\end{tabular}
 \begin{tablenotes}[para,flushleft]
\footnotesize{Note. The table shows the production function coefficients estimated following \citet{Ackerberg_Caves_Frazer_2015} by 2-digit industry based on NACE Rev. 2 classification. The table shows average values and standard deviations in parentheses.}
\end{tablenotes}
\end{threeparttable}
\end{table}


\setcounter{table}{0}
\renewcommand{\thetable}{B\arabic{table}}
\setcounter{figure}{0}
\renewcommand{\thefigure}{B\arabic{figure}}

\section*{Appendix B: Tables and graphs}

\begin{table}[htbp]
\footnotesize
\caption {Variables' description}\label{tab:description}
\centering
 \resizebox{1.0\textwidth}{!}{%
\begin{threeparttable}
\begin{tabular}{llrr}
\hline
\textbf{Variables} & \textbf{Description}                                                                                                   & \textbf{Mean} & \textbf{St. Deviation} \\ \hline
Markup             & estimated following \citet{DeLoecker_Warzynski_2012} & 1.66          & 0.48                   \\
Sales             & as from original financial information & 8,701,449          &  151,000,000                  \\
ROI            & return on investment: profits on fixed assets & 11.08          &  1,752                  \\
Capital intensity         & fixed assets per employee & 61,883
          &   364,787                 \\
TFP        & estimated following \citet{Ackerberg_Caves_Frazer_2015} & 9.76          &  1.19                  \\
Fixed assets        & as from original financial information &  4,345,725          &  138,000,000     \\
Value added       & as from original financial information & 2,733,870          &  51,100,000                 \\
Number of employees       & as from original financial information & 42          &  410                \\
Market Share       & firm's revenues over total by country-sector-year   & 0.009        & 0.008                  \\
Variable costs       & sum of costs of materials and employees   & 6,226,557       & 118,000,000             \\ \hline
\end{tabular}
 \begin{tablenotes}[para,flushleft]
\footnotesize{Note. The table provides a description and summary of the statistics of the variables used in the analysis.}
\end{tablenotes}
\end{threeparttable}
}
\end{table}

\begin{figure}[H]
\caption{Sales-weighted average markup in the European Union, 2007-2021}
\begin{minipage}{1\textwidth}
\centering
\includegraphics[scale=0.12]{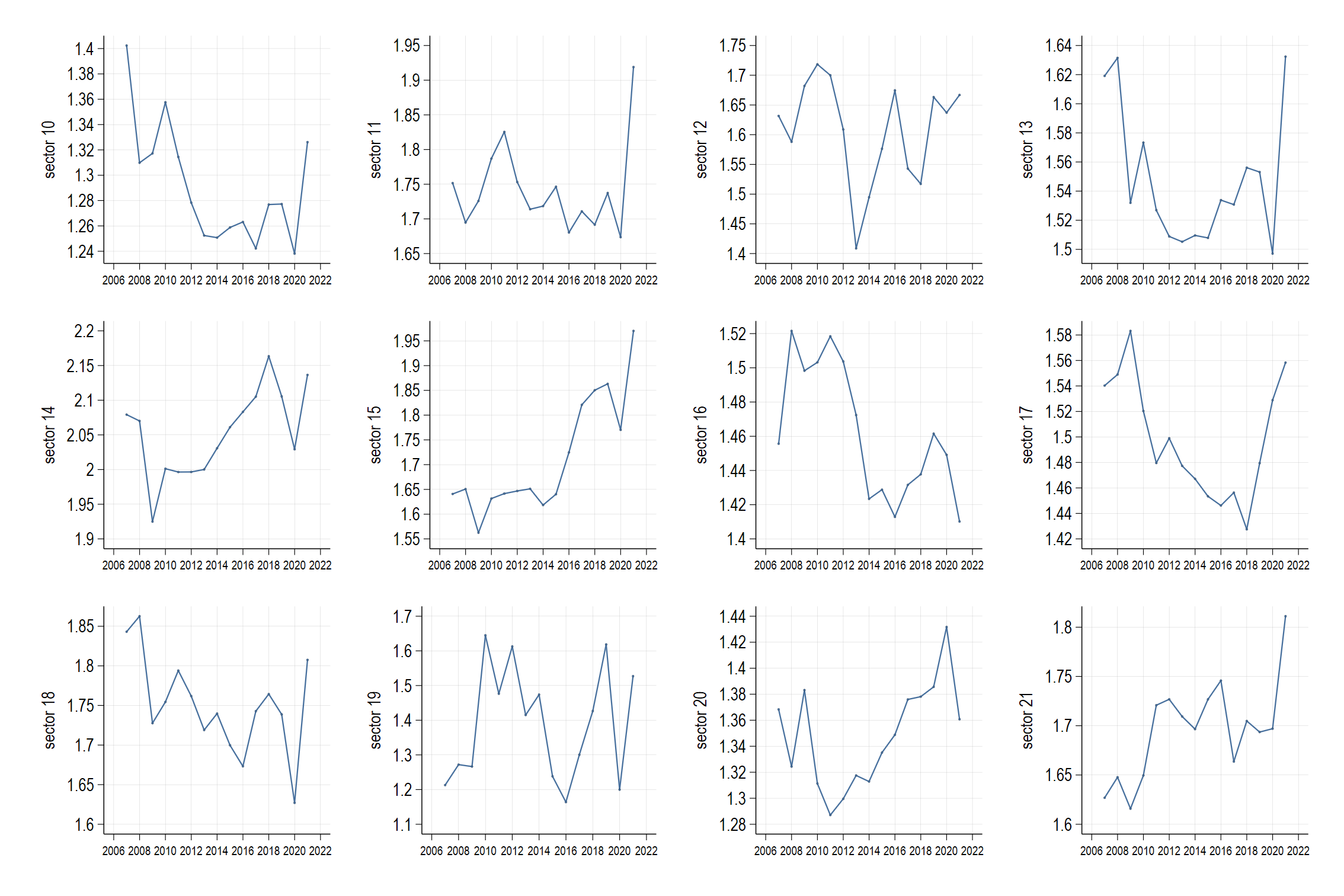}
\includegraphics[scale=0.12]{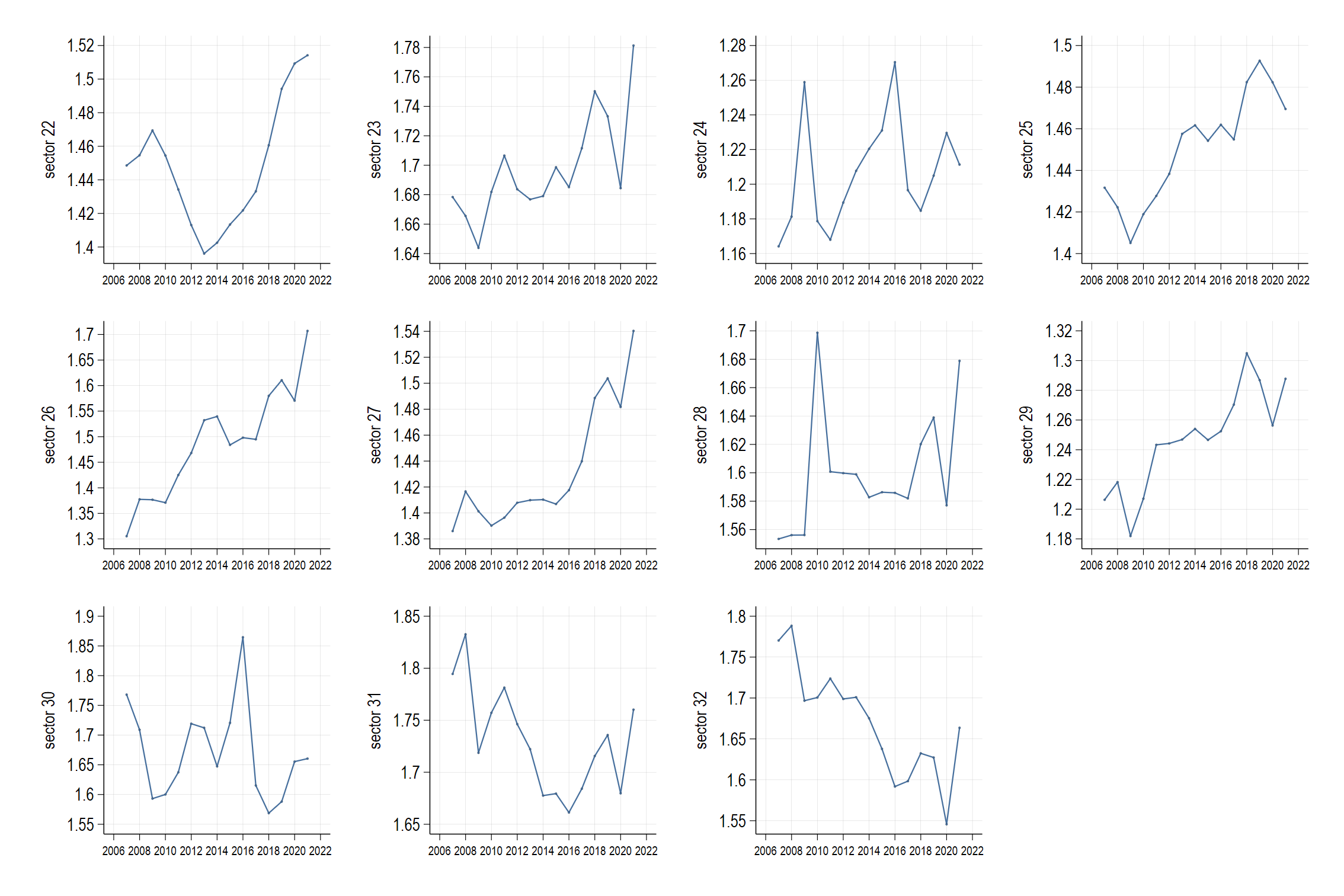}\\
\justifying
\footnotesize{Note. The figure reports the sales-weighted average markup for 2-digit NACE rev. 2 manufacturing industries. Markups are estimated following \citet{DeLoecker_Warzynski_2012}.}
\end{minipage}
\label{fig:fig3}
\end{figure}

\begin{table}[htbp]
\caption{Time coverage of takeovers}\label{tab:year_distribution}
\centering
\small
\begin{threeparttable}
\begin{tabular}{cc}
\multicolumn{1}{l}{Year of acquisition} & N. of acquisitions \\ \hline
2009                                      & 381 \\
2011                                      & 704                            \\
2013                                      & 1,023                            \\
2015                                      & 1,190                           \\
2017                                      & 1,184                           \\
\hline
\multicolumn{1}{c}{Total}                 & 4,482                        
\end{tabular}
 \begin{tablenotes}[para,flushleft]
\footnotesize{Note. The table shows the number of acquisitions per release of the ownership dataset.}
\end{tablenotes}
\end{threeparttable}
\end{table}


\begin{table}[htbp]
\footnotesize
\caption {Industry coverage of firms' acquisitions}\label{tab:sector_distribution}
\centering
\begin{threeparttable}
\begin{tabular}{clr} 
\toprule
NACE &\multicolumn{1}{l}{Industry description}                                                                                                                                 & \multicolumn{1}{c}{N. of acquisitions}  \\ 
\midrule

10	&	Manufacture	of food products	&	478	\\
11	&	Manufacture	of beverages	&	132	\\
12	&	Manufacture	of tobacco products	&	4	\\
13	&	Manufacture	of textiles	&	154	\\
14	&	Manufacture	of wearing apparel	&	87	\\
15	&	Manufacture	of leather and related products	&	56	\\
16	&	Manufacture	of wood and of products of wood and cork	&	132	\\
17	&	Manufacture	of paper and paper products	&	111	\\
18	&	Printing and	 reproduction of recorded media	&	98	\\
19	&	Manufacture	of coke and refined petroleum products	&	9	\\
20	&	Manufacture	of chemicals and chemical products	&	338	\\
21	&	Manufacture	of basic pharmaceutical products	&	84	\\
22	&	Manufacture	of rubber and plastic products	&	310	\\
23	&	Manufacture	of other non-metallic mineral products	&	227	\\
24	&	Manufacture	of basic metals	&	160	\\
25	&	Manufacture	of fabricated metal products	&	706	\\
26	&	Manufacture	of computer, electronic and optical products	&	179	\\
27	&	Manufacture	of electrical equipment	&	208	\\
28	&	Manufacture	of machinery and equipment n.e.c.	&	600 \\
29	&	Manufacture	of motor vehicles, trailers and semi-trailers	&	150	\\
30	&	Manufacture	of other transport equipment	&	51	\\
31	&	Manufacture	of furniture	&	102	\\
32	&	Other manufacturing		&	106	\\ \hline
Total & & 4,482 \\
\bottomrule
\end{tabular}
 \begin{tablenotes}[para,flushleft]
\footnotesize{Note. The table shows the number of acquisitions per 2-digit industry based on NACE Rev. 2 classification.}
\end{tablenotes}
\end{threeparttable}
\end{table}

\begin{table}[htbp]
\footnotesize
\caption {Correlation matrix}\label{tab:corr}
\centering
 \resizebox{0.9\textwidth}{!}{%
\begin{threeparttable}
\begin{tabular}{lrrrrrrrrrrr}
\hline
 	&	markup	&	sales	&	ROI	&	\makecell[c]{capital \\ intensity}	& TFP	&	\makecell[c]{fixed \\ assets}	&	\makecell[c]{value \\ added}	&	\makecell[c]{num. of \\ employees} 	&	\makecell[c]{market \\ shares}	&	\makecell[c]{variable \\ costs}	\\
markup	&	1		\\
sales	&	-0.0275*	&	1		\\
ROI	&	-0.0012	&	-0.0001	&	1	\\
capital intensity	&	-0.0512*	&	0.0472*	&	-0.0013	&	1		\\
TFP	&	0.1173*	&	-0.0139*	&	0.0024*	&	-0.0781*	&	1		\\
fixed assets	&	-0.0064*	&	0.8514*	&	-0.0001	&	0.0746*	&	-0.0082*	&	1		\\
value added	&	-0.0154*	&	0.9137*	&	-0.0001	&	0.0460*	&	-0.0101*	&	0.8479*	&	1		\\
num. of employees	&	-0.0198*	&	0.8635*	&	-0.0001	&	0.0274*	&	-0.0119*	&	0.7317*	&	0.8713*	&	1		\\
market share	&	-0.0331*	&	0.2267*	&	-0.0003	&	0.0538*	&	-0.0111*	&	0.1516*	&	0.2310*	&	0.2565*	&	1		\\
var. cost	&	-0.0275*	&	0.9942*	&	-0.0001	&	0.0429*	&	-0.0139*	&	0.8435*	&	0.8971*	&	0.8491*	&	0.2189*	&	1	\\
            \\ \hline
\end{tabular}
  \begin{tablenotes}[para,flushleft]
\footnotesize{Note. The table shows pairwise correlations of variables for treated and untreated firms included in the sample.}
\end{tablenotes}
\end{threeparttable}
}
\end{table}
\begin{table}[htpb]
\caption{Test for parallel trends}
\label{tab:parallel_trend}
\centering
\resizebox{\textwidth}{!}{%
\begin{threeparttable}
\begin{tabular}{lcccccr}
\toprule
\multicolumn{7}{c}{Markup} \\ 
 & Coefficient & Std. err. & z & P$>|z|$ & \multicolumn{2}{c}{[95\% conf. interval]} \\ \hline
Pre treatment avg & 0.007 & 0.004 & 1.64 & 0.100 & -0.001 & 0.015 \\
Post treatment avg & -0.014 & 0.004 & -3.27 & 0.001 & -0.023 & -0.006 \\
\multicolumn{3}{l}{Pretrend Test: H0 All Pre-treatment are equal to 0} & $\chi^2$(25) = 44.225 & p-value = 0.010 &  &  \\ \hline
\multicolumn{7}{c}{Market share} \\ 
 & Coefficient & Std. err. & z & P$>|z|$ & \multicolumn{2}{c}{[95\% conf. interval]} \\ \hline
Pre treatment avg & -0.005 & 0.019 & -0.24 & 0.807 & -0.041 & 0.032 \\
Post treatment avg & 0.055 & 0.022 & 2.51 & 0.012 & 0.012 & 0.098 \\
\multicolumn{3}{l}{Pretrend Test: H0 All Pre-treatment are equal to 0} & $\chi^2$(25) = 20.318 & p-value = 0.730 &  &  \\ \hline
\multicolumn{7}{c}{Sales} \\ 
 & Coefficient & Std. err. & z & P$>|z|$ & \multicolumn{2}{c}{[95\% conf. interval]} \\ \hline
Pre treatment avg & -0.010 & 0.014 & -0.71 & 0.475 & -0.038 & 0.018 \\
Post treatment avg & 0.048 & 0.014 & 3.37 & 0.001 & 0.020 & 0.076 \\
\multicolumn{3}{l}{Pretrend Test: H0 All Pre-treatment are equal to 0} & $\chi^2$(25) = 34.034 & p-value = 0.107 &  &  \\ \hline
\multicolumn{7}{c}{Variable cost} \\ 
 & Coefficient & Std. err. & z & P$>|z|$ & \multicolumn{2}{c}{[95\% conf. interval]} \\ \hline
Pre treatment avg & -0.008 & 0.016 & -0.51 & 0.612 & -0.039 & 0.023 \\
Post treatment avg & 0.051 & 0.015 & 3.41 & 0.001 & 0.022 & 0.080 \\
\multicolumn{3}{l}{Pretrend Test: H0 All Pre-treatment are equal to 0} & $\chi^2$(25) = 33.015 & p-value = 0.131 &  &  \\ \hline

\multicolumn{7}{c}{ROI} \\ 
 & Coefficient & Std. err. & z & P$>|z|$ & \multicolumn{2}{c}{[95\% conf. interval]} \\ \hline
Pre treatment avg & -0.014 & 0.021 & -0.65 & 0.518 & -0.056 & 0.028 \\
Post treatment avg & 0.060 & 0.017 & 3.60 & 0.000 & 0.027 & 0.093 \\
\multicolumn{3}{l}{Pretrend Test: H0 All Pre-treatment are equal to 0} & $\chi^2$(20) = 18.237 & p-value = 0.572 &  &  \\ \hline
\multicolumn{7}{c}{Capital intensity} \\ 
 & Coefficient & Std. err. & z & P$>|z|$ & \multicolumn{2}{c}{[95\% conf. interval]} \\ \hline
Pre treatment avg & 0.020 & 0.029 & 0.69 & 0.490 & -0.037 & 0.078 \\
Post treatment avg & -0.139 & 0.027 & -5.08 & 0.000 & -0.193 & -0.086 \\
\multicolumn{3}{l}{Pretrend Test: H0 All Pre-treatment are equal to 0} & $\chi^2$(25) = 17.255 & p-value = 0.872 &  &  \\ \hline
\end{tabular}
\begin{tablenotes}
\item Note: The table reports the estimated average pre-treatment and post-treatment effects from the event-study regression shown in Figure \ref{fig:event}, for outcome variables where we find significant effects. For each outcome, we also report a pretrend test, where the null hypothesis (H0) is that all pre-treatment coefficients are jointly equal to zero, representing a particularly strict version of the parallel trends assumption. For markups we cannot reject the null hypothesis because the p-value of the test is significant, but this is the result of a single pre-treatment period which is statistically significant, while the overall pattern of pre-treatment coefficients remains stable as confirmed by the pre-treatment average which is statistically not significant.
\end{tablenotes}
\end{threeparttable}}
\end{table}

\begin{table}[]
\caption {Average treatment effects on the treated (ATET) on markups: least-squares production function} \label{tab:ols} 
\centering
\footnotesize
\resizebox{0.6\textwidth}{!}{%
\begin{tabular}{lcc} \hline
 & (1)  & (2)\\
Variables & Markup - baseline &  Markup - vertical\\ \hline
ATET &  -0.002 &  -0.005*  \\
 & (0.002) & (0.003) \\ \hline
Obs. & 3,707,371 & 3,675,182 \\
\hline
\end{tabular}}\\
\justifying
\footnotesize{Note: The table shows results after the doubly robust estimator proposed by \citet{Callaway_Santanna_2021}, using never-treated and not-yet-treated units in the control group. We test the sensitivity of the results on markups by estimating a production function with simple least squares. Standard errors clustered at the firm level are reported in parentheses, and significance levels are *** p$<$0.01 ** p$<$0.05 * p$<$0.1.}
\end{table}

\begin{table}[!ht]
    \centering
    \caption{Balancing properties for matching takeovers}\label{fig:balancing properties}
    \resizebox{1\textwidth}{!}{%
    \begin{tabular}{lcccccc}
    \hline
         Horizontal integrations & ~ & ~ &  & ~ & ~ & ~ \\ \hline
        Variable & Mean treated & Mean control & \% bias & t & $p>|t|$ & V(T)/V(C) \\ 
        (log of) capital intensity & 10.544 & 10.570 & -1.70 & -1.450 & 0.148 & 1.08 \\ 
        (log of) firm age & 3.306 & 3.321 & -2.60 & -2.05 & 0.040 & 0.77* \\ 
        (log of) firm size & 3.817 & 3.812 & 0.40 & 0.30 & 0.767 & 0.90 \\ 
        (log of) TFP & 9.782 & 9.818 & -3.00 & -2.31 & 0.121 & 0.96 \\ 
        ~ & ~ & ~ & ~ & ~ & ~ & ~ \\ \hline
         Vertical integrations & ~ & ~ &  & ~ & ~ & ~ \\ \hline
        Variable & Mean treated & Mean control & \% bias & t & $p>|t|$ & V(T)/V(C) \\ 
        (log of) capital intensity & 10.648 & 10.655 & -0.50 & -0.57 & 0.568 & 1.07 \\ 
        (log of) firm age & 3.375 & 3.382 & -1.20 & -1.40 & 0.162 & 0.76* \\ 
        (log of) firm size & 3.943 & 3.957 & -1.00 & -1.09 & 0.275 & 0.83 \\ 
        (log of) TFP & 9.710 & 9.727 & -1.50 & -1.63 & 0.104 & 1.03 \\ \hline
    \end{tabular}}\\
\justifying
\footnotesize{Note: The table shows balancing properties after the propensity score matching presented in Table \ref{tab:psm}. Indicator variables like industry and country are not included. * if variance ratio outside the interval $[0.80; 1.20]$}
\end{table}

\begin{table}[htbp]
\caption {Sample coverage by technology intensity}\label{tab:tech_distribution} 
\centering
\begin{threeparttable}
\begin{tabular}{lcc}
Technological Intensity & Frequency & \%    \\ 
\cmidrule[\heavyrulewidth]{1-3}
Low tech                         & 268,372           & 46\%               \\ 
Medium-low tech                  & 195,250

            & 34\%
                \\
Medium-high tech                 & 97,287

            & 17\%               \\
High tech                        & 17,963

             & 3\%               \\
\cmidrule[\heavyrulewidth]{1-3}
\end{tabular}
\begin{tablenotes}[para,flushleft]
 \footnotesize{Note. The table represents sample coverage by technology intensity based on firms' industrial affiliations, as per Eurostat classification.}
\end{tablenotes}
\end{threeparttable}
\end{table}


\end{document}